\documentclass[lettersize,journal]{IEEEtran}
\usepackage{amsmath,amsfonts}
\usepackage{algorithmic}
\usepackage{algorithm}
\usepackage{array}
\usepackage[caption=false,font=normalsize,labelfont=sf,textfont=sf]{subfig}
\usepackage{textcomp}
\usepackage{stfloats}
\usepackage{url}
\usepackage{verbatim}
\usepackage{graphicx}
\usepackage{cite}
\usepackage{booktabs} 
\usepackage{caption,subcaption}

\usepackage{wrapfig}
\usepackage{multicol}
\usepackage{color}
\usepackage{tabularx}
\usepackage{subcaption}

\usepackage{caption}
\usepackage{svg}
\usepackage{titlesec}

\begin{document}
\title{Graph Attention-based Reinforcement Learning for Trajectory Design and Resource Assignment in Multi-UAV Assisted Communication}

\author{Zikai Feng, ~\IEEEmembership{Student Member,~IEEE}, Di Wu, ~\IEEEmembership{Member,~IEEE}, Mengxing Huang, ~\IEEEmembership{Member,~IEEE}, Chau Yuen, ~\IEEEmembership{Fellow,~IEEE}
\thanks{* Corresponding author: Mengxing Huang and Chau Yuen.
}

\thanks{This work is supported by the Scientific Research Fund Project of Hainan University (KYQD(ZR)-21007), the Natural Science Foundation of Hainan Province (621QN212), and the National Natural Science Foundation of China (62062030). This work is also supported under the State Scholarship Fund, China Scholarship Council (202207565036).
}

\thanks{Zikai Feng, Di Wu, and Mengxing Huang are with the School of Information and Communication Engineering, Hainan University, and State Key Laboratory of Marine Resource Utilization in South China Sea, Haikou, 570228, China (13523011824@163.com, hainuwudi@163.com, and huangmx09@163.com).

Zikai Feng is also with the Engineering Product Development Pillar, Singapore University of Technology and Design, 487372, Singapore.

Di Wu is also with the Department of Automation, Shanghai Jiao Tong University, Shanghai, 200240, China, and PRISMA Lab, University of Naples Federico II, Naples, 80125, Italy.

Chau Yuen is with the School of Electrical and Electronic Engineering, Nanyang Technological University, 639798, Singapore (e-mail: chau.yuen@ntu.edu.sg).}
}

\markboth{}%
{Shell \MakeLowercase{\textit{et al.}}: A Sample Article Using IEEEtran.cls for IEEE Journals}

\maketitle
\begin{abstract}
In the multiple unmanned aerial vehicle (UAV)-assisted downlink communication, it is challenging for UAV base stations (UAV BSs) to realize trajectory design and resource assignment in unknown environments. The cooperation and competition between UAV BSs in the communication network leads to a Markov game problem. Multi-agent reinforcement learning is a significant solution for the above decision-making. However, there are still many common issues, such as the instability of the system and low utilization of historical data, that limit its application. In this paper, a novel graph-attention multi-agent trust region (GA-MATR) reinforcement learning framework is proposed to solve the multi-UAV assisted communication problem. Graph recurrent network is introduced to process and analyze complex topology of the communication network, so as to extract useful information and patterns from observational information. The attention mechanism provides additional weighting for conveyed information, so that the critic network can accurately evaluate the value of behavior for UAV BSs. This provides more reliable feedback signals and helps the actor network update the strategy more effectively. Ablation simulations indicate that the proposed approach attains improved convergence over the baselines. UAV BSs learn the optimal communication strategies to achieve their maximum cumulative rewards. Additionally, multi-agent trust region method with monotonic convergence provides an estimated Nash equilibrium for the multi-UAV assisted communication Markov game. 
\end{abstract}


\section{Introduction}
\IEEEPARstart{I}{N} modern communication, 5G mobile network is being implemented globally. Compared to its previous iteration, the latest version brings substantial enhancements on mobile devices \cite{1}. However, due to the fast growth of emerging terminals, ground backbone networks are facing serious challenges of data congestion \cite{2} \cite{3}. On account of geographical environment, some outlying regions still lack sufficient wireless coverage. Traditional ground communication networks are increasingly difficult to meet the demand for high-quality services in wireless communication. Therefore, research must be dedicated to the new generation of 6G communication \cite{4}.

The integration of air-space-ground communication is a significant direction of 6G, which provides a solution for the current insufficient wireless communication coverage \cite{5}. One of these technologies is the widespread use of unmanned aerial vehicle (UAV) \cite{6}\cite{7}. 

UAVs hold tremendous promise due to affordability, strong flexibility, and easy deployment. In the field of wireless communication, they often play roles of aerial access nodes to support the downlink communication \cite{8}. In remote and disaster-prone areas, UAVs provide new ideas for inadequate or ineffective infrastructure \cite{9}; in densely populated and heavily trafficked hotspots, UAVs are utilized to enhance wireless coverage and increase network throughput. Additionally, they can also work in traffic control \cite{10}, industrial Internet of Things \cite{11}, emergency rescue \cite{12}, and other fields \cite{13}\cite{14}.

In order to fully leverage the advantages of drones in communication, researchers have invested a significant amount of interest in the design of UAV-assisted communication systems, such as resource deployment \cite{15}\cite{16}\cite{17}\cite{18}, trajectory optimization \cite{19}\cite{20}, the joint power and trajectory optimization \cite{21}\cite{22}, as well as its further combination with other wireless technologies \cite{23}\cite{24}. For example, non-convex optimization theory is used in \cite{25} to describe UAV based trajectory optimization and resource allocation, and a selectable optimized method is proposed to solve the lower bound problem. It has a better efficiency than the upper bound methods. In \cite{26}, a block coordinate descent method was proposed to decompose non-convex problems such as task allocation, resource scheduling, and trajectory planning of UAVs communication. Lagrangian duality and convex approximation are used to deal with the obtained two sub-problems.

Unfortunately, most of the related work mainly focus on the design and optimization of deterministic systems. In these work, the information of the system, such as the parameters of channels, are known \cite{27}. When it comes to the dynamics and unknowns of communication scenarios under real-world condition, it may not be practical to simply use the traditional optimization methods.

With the realistic demands of high dimensional, nonlinear, and unknown environments in communication systems, the traditional optimization method is becoming more and more difficult to meet the decision-making \cite{28}. It is necessary to use machine learning methods to explore new paths. Entities with learning ability can better cope with the dynamic changes of the communication network. Among the most encouraging solutions, reinforcement learning is a promising one \cite{29}.

The core idea of reinforcement learning is that the agent optimize its strategy by learning interactive data with the environment, thereby obtaining the maximum cumulative reward \cite{30}. Recently, the success of single-agent reinforcement learning has spread to multi-agent scenarios \cite{31}\cite{32}. For instance, multi-agent deep deterministic policy gradient algorithm overcomes the obstacle of policy instability caused by dynamic environmental \cite{33}. Similarly, multi-agent proximal policy optimization algorithm extends the single-agent policy gradient methods to the field of multi-agent systems and inherits its excellent convergence performance\cite{34}.

Some research have already focused on the multi-agent reinforcement learning (MARL) based communication networks \cite{35}. In \cite{36}, a combination of game theory and MARL is utilized to facilitate route mapping for UAV BSs. It introduces low computational complexity methods and distributed characteristics, and designs a dynamic multi-target sensing framework for unmanned aerial vehicles. The multi-agent actor-critic technique is employed in \cite{37} to address the power assignment within UAV-enhanced mobile networks, aiming to maximize the capacity.

However, the above MARL algorithms share certain common problems, such as the instability of the system and low utilization of historical data. This is because, in multi-agent systems, the relationships and topology between agents are usually dynamically changing. There are rich interactions and dependencies between agents, as well as traditional optimization methods cannot directly handle these structured data. Therefore, how to effectively capture the complex interaction relationships between agents and aggregate global information in the topology network, so that each agent can obtain a comprehensive view of the entire environment, is crucial for all agents. Besides, in large-scale communication scenarios, the dynamic environment information and the interaction information between agents are exponential. It is also a challenge for agents to extract the most important information from massive information and eliminate useless information.

Motivated by the above analysis, a novel graph-attention multi-agent trust region reinforcement learning framework, GA-MATR, is proposed in this paper to execute the trajectory design and resource assignment in multi-UAV assisted communication system. Because of the cooperation and competition between UAV base stations (UAV BSs), the downlink communication system with multiple UAV BSs and ground users (GUs) is modeled as a Markov game. After that, graph recurrent neural network is introduced to analyze and process complex communication topology data and extract useful information and patterns. The attention mechanism can provide additional information transmission and weighting, so that the critic network can more accurately evaluate the behavior value for UAV BSs. This provides more reliable feedback signals and helping the actor network to update the strategy more effectively. Besides, the standard deviation regularization is introduced to avoid significant resource allocation gaps between the paired GUs, so as to ensure the fairness in resource allocation. UAV BSs in the communication network employ the enhanced method to optimize their own communication strategy, so as to maximize the cumulative rewards. In addition, with monotonic improvement guarantee of the MARL framework, the joint optimal strategies of UAV BSs gradually converge. This provides an approximate Nash equilibrium solution for the UAV-assisted communication Markov game.  

The main contributions of this work are as bellow:

1) A novel graph-attention multi-agent trust region reinforcement learning framework is proposed to address the trajectory design and resource assignment in UAV-assisted communication. Graph neural network is introduced to process and analyze complex topology and extract useful information from communication data. The attention mechanism provides additional weighting for transferred data, so that the critic network can more accurately evaluate the value of behavior for UAV BSs. It provides more reliable feedback signals and helps the actor network update parameters more effectively.

2) The standard deviation regularization is introduced to avoid significant resource allocation gaps between GUs served by the same UAV BS. This ensures the fairness in resource allocation for UAV BSs.

3) Game theory is employed to model the UAV-assisted communication system, and the downlink communication system with multiple UAV BSs is stated as a Markov game. With monotonic convergence guarantee of multi-agent trust region approach, the joint optimal polices of UAV BSs gradually converge. This provides an approximate solution of Nash equilibrium.

In this paper, the system representation and problem description are described in Section 2. The graph-attention multi-agent trust region framework for the multi-UAV assisted communication Markov game can be found in Section 3. The ablation simulations are analyzed in Section 4. Section 5 summarizes this paper.

\section{System Modeling and Problem Description}
This paper considers a multi-UAV assisted communication system in the \emph{D} region, $D \subset {R^3}$. In this system, \emph{M} UAV BSs provide wireless communication services for \emph{N} GUs. Define the set of UAV BSs as 
$M = \{ 1,2,...,M\}$, and define the set of GUs as $N = \{ 1,2,...,N\}$. The schematic diagram of UAV-assisted communication system is shown as Figure 1.

In this paper, all GUs move randomly on the two-dimensional (2D) ground, and each UAV BS flies in the three-dimensional (3D) space and gives wireless services for GUs using frequency division multiple-plexing (FDM). At each time slot, UAV BSs will first independently and sequentially decide to pairing with GUs, and broadcast the pairing decision and position information to other UAV BSs. After that, they will design trajectories based on the positions of the paired GUs, allocate power and bandwidth resources in real-time. In order to reduce interference and improve the quality of communication, each GU will only be paired with one UAV BS. This can also ensure that every GU can receive assisted wireless communication services. In this paper, to avoid wasting resources, we assume each UAV BS will serve at least one GU. Multi-beamforming is used in power allocation. The total power and bandwidth resources for each UAV BS are denoted as $P_{\text {total }}$ and $B_{\text {total }}$ respectively in this paper, which are fixed and the same. Besides, each UAV BS has four different resource allocation options to power and bandwidth respectively at each time slot. Each UAV BS aims to maximize the mean throughput of the paired GUs, as well as ensuring fairness.
\begin{figure}[htbp]
\centerline{\includegraphics[width=9cm]{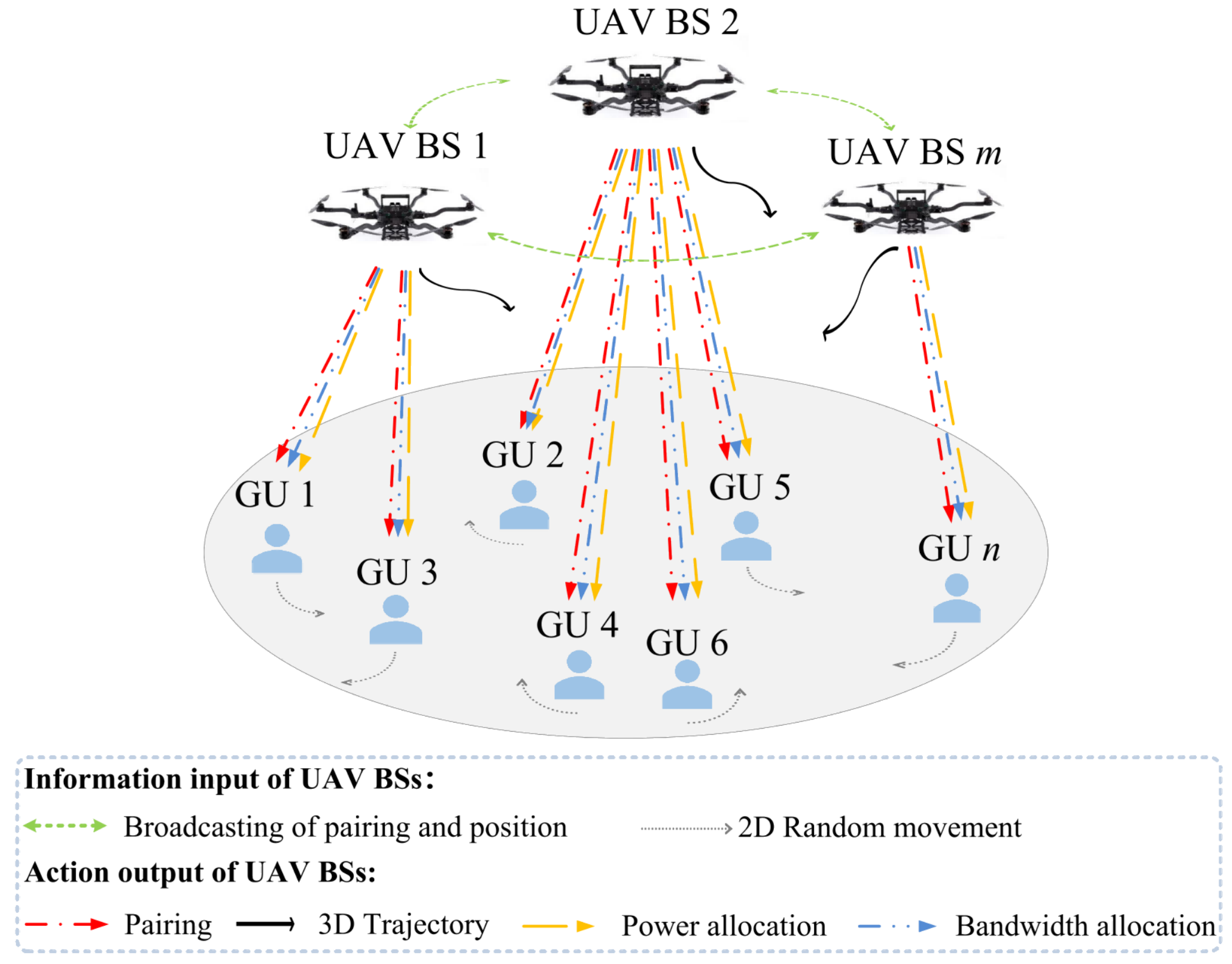}}
\caption{Schematic diagram of UAV-assisted communication system}
\label{fig}
\end{figure}
\subsection{\textbf{Air-ground communication Model}}
Statistical probability-based air-ground communication models have been applied in many works. Specifically, the down-link channel can be separated into Line of Sight (LoS) and Non-Line of Sight (NLoS) \cite{38}. 

In this work, the coordinates of the UAV BS \emph{m}, and the GU \emph{n} are expressed as $P_{os}^m = ({x_m},{y_m},{z_m})$ and $P_{os}^n = ({x_n},{y_n},0)$ respectively, where $P_{o s}^n=\left(x_n, y_n, 0\right)$, $x_m, x_n \in(-100,100)$, and $z_m \in(0,100)$. Due to the limitations of the flight altitude of UAV BSs, the downlink channel is typically LoS. Therefore, at time \emph{t}, the total path loss from UAV BS \emph{m} to GU \emph{n} in the LoS link can be described as
\begin{equation}
P L_{m, n}(t)=l_{m, n}^{L o S}(t)+\varsigma L o S
\end{equation}
where $l_{m, n}^{L o S}(t)$ is the instant path loss from UAV BS \emph{m} to GU \emph{n} on the LoS link, and $\varsigma \operatorname{LoS}$ is the additional losses under the free space transmission.
\begin{equation}
l_{m, n}^{L o S}(t)=20 \lg \left(\frac{4 \pi f_c d_{m, n}(t)}{c}\right)
\end{equation}
where ${f_c}$ is the carrier frequency; ${d_{m,n(t)}} = h/\sin ({\varphi _{m,n}}(t))$ is the straight-line distance from UAV BS \emph{m} to GU \emph{n}, and \emph{c} is the speed of light. ${P_{m,n}}(t)$ is the transmit power allocated by UAV BS \emph{m} to GU \emph{n} at time \emph{t}.

Then, the received signal power available at GU \emph{n} is $ P_{m,n}^{rec}(t) = {P_{m,n}}(t) - P{L_{m,n}}(t) $.

Considering the flight safety of UAV BSs, their minimum flight altitude is set to 5 meters. In Cartesian coordinate system, the velocity of UAV BS \emph{m} is ${v_m} = ({v_m}\mathord{\buildrel{\lower3pt\hbox{$\scriptscriptstyle\rightharpoonup$}} 
\over i}  + {v_m}\mathord{\buildrel{\lower3pt\hbox{$\scriptscriptstyle\rightharpoonup$}} 
\over j}  + {v_m}\mathord{\buildrel{\lower3pt\hbox{$\scriptscriptstyle\rightharpoonup$}} 
\over k} )$.

The location update equation of UAV BS \emph{m} is:
\begin{eqnarray}
P_{os}^m(t + 1) = P_{os}^m(t) + {\rm{v}}_m^t * \Delta t, \,\Delta t = 0.1s 
\end{eqnarray}

GUs move randomly at a speed lower than UAV BSs. At every time slot, the GU will be paired by one UAV BS and be provided communication services. In addition, each UAV BS must serve at least one GU at each time slot.

Then, the connection relationship between UAV BS \emph{m} and GU \emph{n} is represented by binary variables ${\sigma _{m,n}}(t) \in \{ 0,1\} $. When GU \emph{n} is paired by UAV BS \emph{m}, ${\sigma _{m,n}}(t)$ is set to 1, otherwise it is set to 0.
\begin{eqnarray}
{\sigma _{m,n}}(t) = \left\{ \begin{array}{l}
1{, \, ^{}}i{f^{}}\,G{U^{}}{\,n^{}}\,is\ paired\ b{y^{}}\,UAV BS{^{}}\,m\\
0{,^{}}\,\,otherwise
\end{array} \right.
\end{eqnarray}

In the air-ground communication system, the action ${a_m} \in {A_m}$ of UAV BS \emph{m} include the moving vectors ${v_m}$, the option of whether to pair with every GU $one hot\emph{1}{_m}$, the power allocation options $one hot\emph{2}{_m},$ and the bandwidth allocation options $one hot\emph{3}{_m},$ i.e, ${a_m = \{{v_m}, one hot\emph{1}{_m},one hot\emph{2}{_m},one hot\emph{3}{_m}}\}$. The dimension of $one hot\emph{1}{_m}$ is the number of GUs \emph{N}. 

For UAV BS \emph{m}, there are four power allocation options and four bandwidth allocation options respectively, which are shown as below:

1) Allocate communication resources (power and bandwidth) based on randomly generated proportions. 
\begin{eqnarray}
\begin{array}{l}
{P_{m,n}}(t) = {{\tilde p}_{mn}}(t)*{P_{total}},\mathop {}\limits^{} \\
{{\tilde p}_{mn}}(t) = \left\{ \begin{array}{l}
random(0,1)\mathop {}\limits^{}, \,if\,{\sigma _{m,n}}(t) = 1\\
0\mathop {}\limits^{} \mathop {}\limits^{} \mathop {}\limits^{} \mathop {}\limits^{} \mathop {}\limits^{} \mathop {}\limits^{} \mathop {}\limits^{} \mathop {}\limits^{} \mathop {}\limits^{} \,\,\,\,\,\,\,\,\,\,\,\,\,\,\,\,\,\,\,\,\,\,\,\,, if\,{\sigma _{m,n}}(t) = 0
\end{array} \right.,\\
s.t.\mathop {}\limits^{} \sum\nolimits_{i = 0}^N {{{\tilde p}_{mn}}(t)}  = 1.
\end{array}
\end{eqnarray}
\begin{eqnarray}
\begin{array}{l}
{B_{m,n}}(t) = {{\tilde b}_{mn}}(t)*{B_{total}},\mathop {}\limits^{} \\
{{\tilde b}_{mn}}(t) = \left\{ \begin{array}{l}
random(0,1)\mathop {}\limits^{} \,, if\,{\sigma _{m,n}}(t) = 1\\
0\mathop {}\limits^{} \mathop {}\limits^{} \mathop {}\limits^{} \mathop {}\limits^{} \mathop {}\limits^{} \mathop {}\limits^{} \mathop {}\limits^{} \mathop {}\limits^{} \mathop {}\limits^{} \,\,\,\,\,\,\,\,\,\,\,\,\,\,\,\,\,\,\,\,\,\,\,\,, if\,{\sigma _{m,n}}(t) = 0
\end{array} \right.,\\
s.t.\mathop {}\limits^{} \sum\nolimits_{i = 0}^N {{{\tilde b}_{mn}}(t)}  = 1.
\end{array}
\end{eqnarray}
where ${\tilde p_{mn}}(t)$ is the proportion of power obtained by GU \emph{n} from UAV BS \emph{m} to the total power; ${B_{m,n}}(t)$ is the bandwidth obtained by GU \emph{n} from UAV BS \emph{m}; ${\tilde b_{mn}}(t)$ is the proportion of bandwidth obtained by GU \emph{n} from UAV BS \emph{m} to the total bandwidth.

2) Allocate communication resources evenly based on the total number of paired GUs. 
\begin{eqnarray}
{P_{m,n}}(t) = \frac{1}{{{s_m}(t)}}{P_{total}}
\end{eqnarray}
\begin{eqnarray}
{B_{m,n}}(t) = \frac{1}{{{s_m}(t)}}{B_{total}}
\end{eqnarray}
where ${s_m}(t)$ presents the number of GUs paired by UAV BS \emph{m} at time slot \emph{t}.

3) Fully allocate communication resources based on the distance.
\begin{eqnarray}
{P_{m,n}}(t) = \frac{{{P_{total}}}}{{{d_{mn}}{{(t)}^\alpha }}}
\end{eqnarray}
\begin{eqnarray}
{B_{m,n}}(t) = \frac{{{B_{total}}}}{{{d_{mn}}{{(t)}^\alpha }}}
\end{eqnarray}

4) Allocate communication resources based on the distance between UAV BS \emph{m} and GU \emph{n}, plus the weighted sum of random proportions.
\begin{eqnarray}
{P_{m,n}}(t) = {c_1}*\frac{{{P_{total}}}}{{{d_{mn}}{{(t)}^\alpha }}} + {c_2}*{\tilde p_{mn}}(t),\mathop {}\limits^{} {c_1} + {c_2} = 1
\end{eqnarray}
\begin{eqnarray}
{B_{m,n}}(t) = {c_3}*\frac{{{B_{total}}}}{{{d_{mn}}{{(t)}^\alpha }}} + {c_4}*{\tilde b_{mn}}(t),\mathop {}\limits^{} {c_3} + {c_4} = 1
\end{eqnarray}

For example, if $ N = 6 $ and ${\rm{one}} hot\emph{1}{_m} = [1,0,0,0,1,0]$, UAV BS \emph{m} selects the GU \emph{1} and the GU \emph{5}. If $one hot\emph{2}{_m} = [1,0,0,0]$, UAV BS \emph{m} selects the power allocation method \emph{1}; if $one hot\emph{3}{_m} = [0,1,0,0]$, UAV BS \emph{m} selects the bandwidth allocation method \emph{2}.

Define ${n_0}$ as the power spectral density, and ${b_{m,n}}(t)$ is the bandwidth resources form UAV BS \emph{m} to GU \emph{n}. Based on the Shannon channel capacity theory, the communication rate provided by UAV BS \emph{m} to GU \emph{n} is shown as
\begin{eqnarray}
{c_{m,n}}(t) = {\sigma _{m,n}}(t){B_{m,n}}(t){\log _2}(1 + \frac{{P_{m,n}^{rec}(t)}}{{{n_0}{B_{m,n}}(t)}})
\end{eqnarray}

\subsection{\textbf{Optimization Problem Description}}

In this work, in order to maximize the throughput of the multi-UAV assisted communication system, as well as ensuring the fairness in resource allocation, each UAV BS needs to optimize the pairing strategy firstly, then to optimize the flight trajectory, the power allocation strategy, and the bandwidth allocation strategy. Take ${T_{\max }}$ as the time period of one episode, and take ${t}$ as the time interval for a decision $ t \in T = \{ 1,2,...,{T_{\max }}\} $.

Figure 2 gives the time series decision-making process of the multi-UAV assisted communication network. At each time slot \emph{t}, it is first necessary for each UAV BS to decide to pair which GUs. Specifically, UAV BSs will select to pair with GUs in sequence. In order to reduce interference and improve the quality of communication, each GU will only be paired with one UAV BS. The GU that has been paired with the previous UAV BSs is no longer available to the next UAV BSs, which can avoid repeated pairing. After determining the GUs to be paired, the UAV BS will broadcast the pairing decision and position information to all other UAV BSs. In this paper, we assume each UAV BS will serve at least one GU, so as to avoid wasting resources. After that, each UAV BS will adjust it's own trajectory in real-time based on the current positions of the paired GUs (since GUs are moving as well) to reduce the overall path loss. At the same time, each UAV BS will adjust the power and bandwidth allocation strategy according to the current topology and channel conditions, so that each GU can enjoy the same communication services as much as possible. For every UAV BS, the observation information in each time slot includes the positions of the paired GUs, the positions of other UAV BSs, and pairing information of other UAV BSs.

To ensure fairness in resource allocation for UAV BSs, this paper introduces the index of standard deviation regularization $ {\varepsilon _{m}} $ as an evaluation index for UAV BS \emph{m}
\begin{eqnarray}
{\varepsilon _{m}}  = \sqrt {\frac{{\sum\nolimits_{n = 1}^N {({c_{m,n}} - \overline c )} }}{n}} ,\mathop {}\limits^{} \mathop {}\limits^{} {\overline c_m}  = \frac{1}{N} \sum\nolimits_{n = 1}^N {{c_{m,n}}} ,\,n \in N
\end{eqnarray}
where $ {\overline c_m} $ is the average communication rate of all GUs served by UAV BS \emph{m}. The smaller index $ {\varepsilon_m} $, the higher the fairness of resource allocation for UAV BS \emph{m}, and the smaller the difference in communication rates of GUs.

This paper defines the optimization goal of each UAV BS as:
\begin{eqnarray}
\begin{aligned}
&\text{max} \sum_{t\in T}\sum_{n\in s_{m}(t)}c_{m,n}(t)-\sum_{t\in T}\varepsilon_{m}  \\
&s.t. \sigma_{m,n}(t)\in\{0,1\},n\in N  \\
&\begin{aligned}\,\,\,\,\,\,\sum_{m\in M}\sigma_{m,n}(t)=1,n\in N\end{aligned} \\
&\begin{aligned}\,\,\,\,\,\,\sum_{n\in N}\sigma_{m,n}(t)\geq1,\end{aligned} \\
&\begin{aligned}\,\,\,\,\,\,\sum_{n\in N}\sigma_{m,n}(t)=s_{m}(t),\end{aligned} \\
&\begin{aligned}\,\,\,\,\,\,\sum_{n\in s_{m}\left(t\right)}p_{m,n}(t)=P_{total},\end{aligned} \\
&\begin{aligned}\,\,\,\,\,\,\sum_{n\in s_{m}(t)}b_{m,n}(t)=B_{total},\end{aligned} \\
&\,\,\,\,\,\,b_{m,n}(t)\geq b_{\operatorname*{min}},p_{m,n}(t)\geq p_{\operatorname*{min}},n\in N \\
&\begin{aligned}\,\,\,\,\,\,\forall t\in T,m\in M\end{aligned}
\end{aligned}
\end{eqnarray}
where ${p_{m,n}}(t)$ and ${b_{m,n}}(t)$ are the transmit power and bandwidth resources allocated by UAV BS \emph{m} to GU \emph{n} respectively; ${b_{\min }}$ is the minimum divisible bandwidth; ${p_{\min }}$ is the minimum divisible power.

\begin{figure*}[h]
\centerline{\includegraphics[width=17cm]{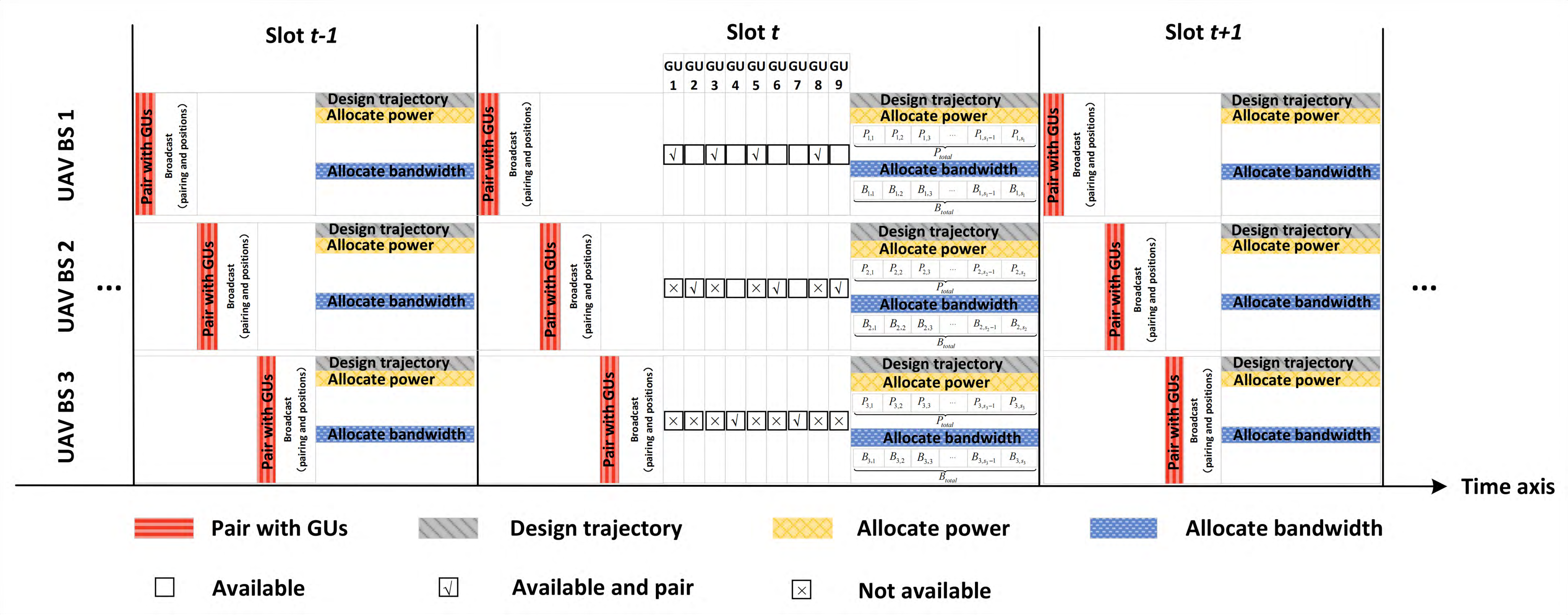}}
\caption{The time series decision-making process with three UAV BSs and nine GUs}
\label{fig}
\end{figure*}

\subsection{\textbf{Multi-UAV Assisted Communication Markov Game}}
Considering the uncertainty of the environment in this paper, intelligent UAV BSs need to make decisions in every time slot, and these decisions will affect the future time slot. Therefore, the Markov framework of sequential decision is adopted.  

1) \textbf{Markov Game}

In a dynamic environment involving multiple UAV BSs, each UAV BS sequentially makes decisions through interactions with the unknown surrounding. Then, the process of joint decision-making involving multiple participants is commonly referred to a stochastic game \cite{39}, alternatively defined a Markov game \cite{40}.

\textbf{Definition 1} Markov Game expands the concept of Markov Decision Process into the multi-agent scenarios \cite{41}. In this work, the problem of multi-UAV assisted communication is described as a tuple of Markov game.

\emph{N}: the quantity of UAV BSs.

\emph{S}: the state set of UAV BSs.

\emph{A}: the joint action set of all UAV BSs, $A: = {A_1} \times  \cdot  \cdot  \cdot  \times {A_N}$.

\emph{P}: $S \times A \to \Delta (S)$: when UAV BSs take joint actions ${\rm{a}} \in A$, the state \emph{s} transfer to the next state \emph{s}’ subject to the probability \emph{P}.

$\gamma  \in [0,1]$ is the discount factor.

For UAV BS \emph{i}:

${o_i}$ is the observation.

${\pi _i}({a_i}|{o_i})$ is the action.

${r_i}:S \times A \to R$ is the reward function, $r_i^t$ is the instant reward.

The UAV BS engage with the environment following this protocol: at time slot \emph{t}, based on the current state ${s_t} \in S$, UAV BS \emph{i} takes an action $a_t^i \in {A^i}$ with its policy ${\pi ^i}( \cdot |{s_t})$. UAV BSs’ joint action ${a_t} = (a_t^1,...,a_t^n) \in A$ corresponds to the joint policy $\pi ( \cdot |{s_t}) = \prod\nolimits_{i = 1}^n {{\pi ^i}({ \cdot ^i}|{s_t})} $. Then, the UAV BS receives a return ${r_t} = r({s_t},{a_t}) \in R$, and moves to the next state ${s_{t + 1}}$ subject to probability $P({s_{t + 1}}|{s_t},{a_t})$. 

The definitions of the state value and the state-action value are established: ${V_\pi }(s) = {{\rm E}_{{a_{0:\infty }} \sim \pi ,{s_{1:\infty }} \sim P}}[\sum\nolimits_{t = 0}^\infty  {{\gamma ^t}} {r_t}|{s_0} = s]$ and ${Q_\pi }(s,a) = {{\rm E}_{{s_{1:\infty }} \sim P,{a_{1:\infty }} \sim \pi }}[\sum\nolimits_{t = 0}^\infty  {{\gamma ^t}} {r_t}|{s_0} = s,{a_0} = a]$. The formulation of the advantage function is expressed as ${A_\pi }(s,a) = {Q_\pi }(s,a) - {V_\pi }(s)$.

The immediate reward for UAV BS \emph{m} is set to:
\begin{eqnarray}
{r_m} = \frac{1}{{{s_m}(t)}}\sum\nolimits_{i = 1}^{{s_m}(t)} {{c_{m,n}}(t)}  - {\varepsilon_m} *\lambda 
\end{eqnarray}
where $\lambda $ is coefficient of the standard deviation regularization index. 

Entities aim to enhance their movement strategies in pursuit of maximizing their overall returns. Furthermore, the description of the optimization model is shown as:
\begin{eqnarray}
\begin{array}{l}
\mathop {\max }\limits_{} {R_m}[{a_m}, m \in M]
\end{array}
\end{eqnarray}
where ${R_m} \! = \!\sum \begin{array}{l}
\infty \\
t = 0
\end{array} \!\!\!{\gamma ^t}{r_m}(t)$ represents the cumulative reward of the UAV BS \emph{m}; 
$\gamma $ is the discount factor. 

2) \textbf{Nash Equilibrium}

Nash Equilibrium (NE) is a key notion in game theory and is first proposed by John Nash in the 1950s \cite{42}. It describes a scenario in which multiple parties participate in decision-making.  

Various approaches exist for addressing Markov games, and Nash equilibrium stands out as one of the most prominent solutions. 

The following gives the definition of NE.

\textbf{Lemma 1}. In Markov game, Nash equilibrium is a set of strategies that meet the following attributes: no participant can independently alter their own strategy to achieve achieve greater profits.

The Markov game discussed in this work is a non-cooperative game. In this non-cooperative game, NE indicates that every participant is rational. In order to distinguish UAV BS \emph{i} and other UAV BSs, the tuple of $( \cdot i, \cdot  - i)$ is introduced to describe the variables. Given the optimal strategies $\pi _{ - i}^*$ of other UAV BSs, every rational UAV BS will not individually change the optimal strategy $\pi _i^*$. In other words, each UAV BS has already made the optimal decision under the strategy of others.

Then, $(\pi _i^*,\pi _{ - i}^*)$ composes a Nash equilibrium for the multi-UAV assisted communication Markov game \emph{G}. Its mathematical form is represented as:
\begin{eqnarray}
{R_i}(\pi _i^*,\pi _{ - i}^*) \ge {R_i}(\pi _i^{},\pi _{ - i}^*),\,i \in N
\end{eqnarray}

\section{Graph Attention-based Trust Region MARL Solution For UAV-Assisted Communication Markov Game}

In this section, we first introduce the proposed graph attention-based trust region MARL algorithm, then analyze the computational complexity of the proposed algorithm and benchmark algorithms, and finally discuss the solution of equilibrium for the UAV-assisted communication Markov game.

\subsection{\textbf{Graph Attention-based Trust Region MARL}}

Trust region reinforcement learning (TRRL) is a trust region based policy optimization algorithm, which aimes to solve non convex, high-dimensional problems \cite{43}. The TRRL method can mathematically prove convergence under certain conditions. For example, the monotonic convergence of Trust Region Policy Optimization (TRPO) reinforcement learning algorithm has been strictly proved in mathematics \cite{44}. Therefore, compared to some traditional gradient optimization methods, TRRL usually has better stability and convergence.

Unfortunately, in terms of multi-agent scenario, the nature of monotonic improvement is no longer simple and applicable, and it mainly faces the challenges of instability, high-dimensional state space and non-stationary.

This is because that policy updates for UAV BSs may lead to rapid changes of the multi-agent environment, leading to instability in the training process. The TRPO algorithm itself is relatively sensitive to stability. In multi-agent environments, due to dynamic changes in the policies of other UAV BSs, the algorithm may be more prone to falling into an unstable state. Besides, the overall state space is often of substantial size. This leads to an explosion in the dimensions of the problem, increasing the difficulty and computational complexity of optimization. In response to these challenges, researchers have proposed many improved and variant TRPO algorithms, such as heterogeneous-agent TRPO \cite{45}, multi agent proximal policy optimization \cite{46}, etc., aimed at enhancing the adaptability and performance of TRPO in multi-agent environments. However, multi-agent reinforcement learning is still an active research field, and more work is needed to solve its unique challenges and problems.

In this paper, based on the stable convergence capability of trust region MARL, we further propose a novel graph attention-based trust region MARL framework, witch can address the instability of the system and low utilization of historical data for the UAV-enhanced communication task. The monotonic improvement property of graph attention trust region MARL have been justified in theory. There is no need for UAV BSs to exchange parameters, and it is also no necessity for limiting assumptions regarding the separability of the value function.

1) \textbf{Trust region reinforcement learning algorithms}

The single-agent trust region method, such as TRPO, was proposed to guarantee the monotonic improvement ${J_\pi }$ in each iteration. The following gives its description.

\textbf{Theorem 1} \cite{47}. Let $\pi $ denotes the present policy and $\bar \pi $ represents the candidate policy. 

There have ${L_\pi }(\bar \pi ) = {J_\pi } + {{\rm E}_{{\rm{s}} \sim \rho ,a \sim \bar \pi }}[{A_\pi }(s,a)]$ and $D_{KL}^{\max}(\pi,\bar\pi) = {\max _s}{D_{KL}}(\pi(\cdot|s),\bar\pi(\cdot |s))$.
Then, the following inequality holds
\begin{eqnarray}
J(\bar \pi ) \ge {L_\pi }(\bar \pi ) - CD_{KL}^{\max }(\pi ,\bar \pi)
\end{eqnarray}
where $C = \frac{{4\gamma {{\max }_{s,a}}|{A_\pi }(s,a)|}}{{{{(1 - \gamma )}^2}}}$.

As the gap between the current strategy $\pi $ and the candidate strategy $\bar \pi $ narrows, UAV BSs that only involve the state distribution of the current strategy ${L_\pi }(\bar \pi )$ will become increasingly accurate estimates of actual performance indicators $J(\bar \pi )$. Based on this, an iterative trust region method is formulated. 

The strategy update formula is as follows:
\begin{eqnarray}
{\pi _{k + 1}} = {\arg _\pi }\max ({L_{{\pi _k}}}(\pi ) - CD_{kl}^{\max }({\pi _k},\pi ))
\end{eqnarray}

This type of update ensures a monotonic enhancement for strategy, i.e., $J({\pi _{k + 1}}) > J({\pi _k})$.

2) \textbf{Trust region-MARL with monotonic convergence}

This work presents a solution for Markov games using MARL, as there exists a natural integration between game theory and MARL. In this section, the trust region algorithm is expand into the field of MARL. 

Acorrding to \cite{45}, the combined advantage can be decoupled as the sum of individual advantages. This decomposition mechanism of advantage values provides key support for ensuring the monotonicity of subsequent sequential policy updates.

Corresponding to Theorem 1, in MARL, with the updates of 
$\pi _{k + 1}^{i{}_m} = \mathop {\arg \max }\limits_{{\pi ^{{i_m}}}} [L_{{\pi _k}}^{{i_{1:m}}}(\pi _{k + 1}^{i{}_{1:m - 1}},{\pi ^{{i_m}}}) - CD_{KL}^{\max }(\pi _k^{{i_m}},{\pi ^{{i_m}}})]$, we have
\begin{eqnarray}
\begin{array}{l}
J({\pi _{k + 1}}) \ge {L_{{\pi _k}}}({\pi _{k + 1}}) - CD_{KL}^{\max }({\pi _k},{\pi _{k + 1}})\\
\,\,\,\,\,\,\,\,\,\,\,\,\,\,\,\,\mathop {}\limits^{} \mathop {}\limits^{} \mathop {}\limits^{} \mathop {}\limits^{} \mathop {}\limits^{}  \ge {L_{{\pi _k}}}({\pi _{k + 1}}) - \sum\limits_{m = 1}^n {CD_{KL}^{\max }(\pi _k^{{i_m}},\pi _{k + 1}^{{i_m}})} \\
\,\,\,\,\,\,\,\,\,\,\,\,\,\,\,\,\mathop {}\limits^{} \mathop {}\limits^{} \mathop {}\limits^{} \mathop {}\limits^{} \mathop {}\limits^{}  = J({\pi _K}) + \sum\limits_{m = 1}^n {(L_{{\pi _k}}^{{i_{1:m}}}(\pi _{k + 1}^{{i_{1:m - 1}}},\pi _{k + 1}^{{i_m}})}  \\
- CD_{KL}^{\max }(\pi _k^{{i_m}},\pi _{k + 1}^{{i_m}}))\\
\,\,\,\,\,\,\,\,\,\,\,\,\,\,\,\,\mathop {}\limits^{} \mathop {}\limits^{} \mathop {}\limits^{} \mathop {}\limits^{} \mathop {}\limits^{}  \ge J({\pi _K}) + \sum\limits_{m = 1}^n {(L_{{\pi _k}}^{{i_{1:m}}}(\pi _{k + 1}^{{i_{1:m - 1}}},\pi _k^{{i_m}})}  \\
- CD_{KL}^{\max }(\pi _k^{{i_m}},\pi _k^{{i_m}}))\\
\,\,\,\,\,\,\,\,\,\,\,\,\,\,\,\,\mathop {}\limits^{} \mathop {}\limits^{} \mathop {}\limits^{} \mathop {}\limits^{} \mathop {}\limits^{}  = J({\pi _K}) + \sum\limits_{m = 1}^n {0 = J({\pi _K})} 
\end{array}
\end{eqnarray}

This proves that trust region achieves monotonic improvement in MARL. With the above theorem, the monotonic improvement property of TRPO has been successfully extended and introduced into MARL.

3) \textbf{Graph attention network (GAN)}

GAN embeds the attention mechanism into the graph neural network \cite{48}. In GAN, an undirected graph ${\rm H} < \eta ,\zeta  > $ contains a vertex $\xi $ for each agent $i = 1,2,...,n$ and a set of undirected edges$\{ i,j\}  \in \zeta $ between connected vertices ${\eta _i}$ and ${\eta _j}$.

Attention mechanism refers to mapping a set of queries and key values to generate an output value. The key value in the query function should be assigned through an attention (weight allocation) mechanism. This can reflect which information needs more attention at a certain moment. The functional expression is provided as follows:
\begin{eqnarray}
Attention(q,\kappa ,\mu ) = soft(\frac{{q{\kappa ^T}}}{{\sqrt {{d_\kappa }} }})\mu 
\end{eqnarray}
where \emph{q} is the query; $\kappa $ and $\mu $ form a key-value pair; ${d_\kappa }$ is the columns of the \emph{q}, ${d_\kappa }$ matrix, i.e., the vector dimension.

GAN integrates the attention mechanism with the graph structure. When calculating the information of each node, additional information of other nodes are introduced. Besides, weights are assigned through an attention network to display the impact of other nodes on their own. The weight expression is:
\begin{eqnarray}
{\varpi _{ij}} = \frac{{\exp (Leaky{\mathop{\rm Re}\nolimits} LU({\varpi ^T}[W{h_i}||W{h_j}]))}}{{\sum\nolimits_{k \in {N_i}} {\exp (Leaky{\mathop{\rm Re}\nolimits} LU({\varpi ^T}[W{h_i}||W{h_j}]))} }}
\end{eqnarray}
where $\varpi $ and \emph{W} are parameters; ${h_i}$ presents eigenvectors of information for node. ${\varpi _{ij}}$ is a weight scalar, which reveals the importance of node ${\eta _i}$ to node ${\eta _j}$.

4) \textbf{Graph attention based trust region MARL framework}

Based on the monotonic improvement capability of trust region MARL, in order to further address the instability and low data utilization caused by dynamic environment, we propose a novel graph attention based trust region MARL framework to coup with the multi-UAV assisted communication.

On one hand, the use of graph recurrent network (GRN) helps to model and handle complex relationships between intelligent UAV BSs. Due to the interaction and dependency relationships between UAV BSs in multi-agent environments, GRN can effectively capture these relationships. By using GRN for representation learning, the observation information of UAV BSs can be mapped to a shared representation space, thereby better expressing the relationships between UAV BSs and GUs. In addition, the graph structure in multi-agent environments is usually dynamically changing with the actions and interactions of the UAV BSs. GRN has the ability to process dynamic graphs and can adaptively update graph structures and node representations, thereby better reflecting changes in states and relationships between UAV BSs and GUs. 

On the other hand, the main idea of combining attention mechanism with MARL is to help critic networks selectively focus on interactions or important features related to the current task. By learning attention weights, critic networks can focus more attention on UAV BSs that contribute to the current decision based on task requirements, thereby improving the accuracy and generalization ability of state or action value functions.
\begin{figure*}
\centerline{\includegraphics[width=15cm]{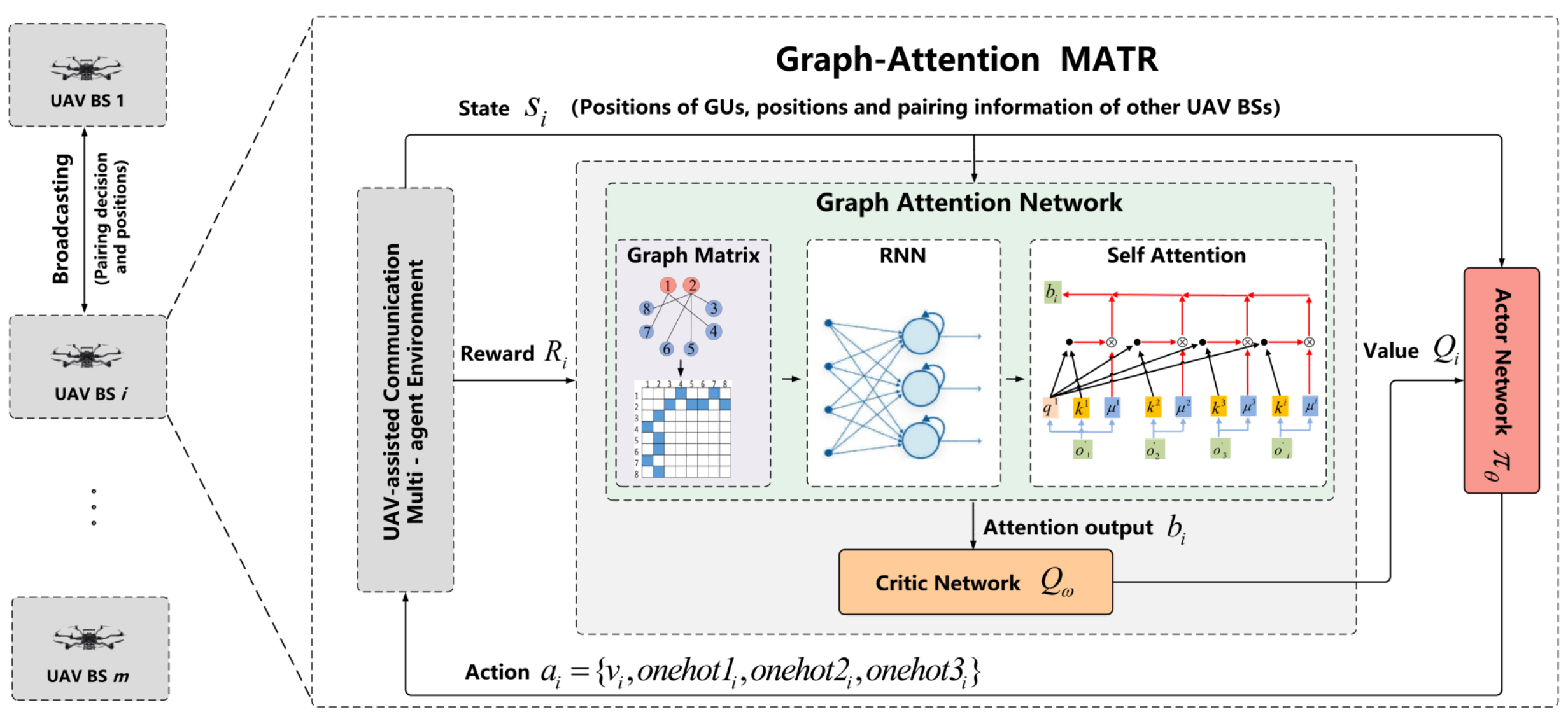}}
\caption{Framework of graph attention-based multi-agent trust region reinforcement
learning
}
\label{fig}
\end{figure*}

\begin{algorithm} 
	\renewcommand{\algorithmicrequire}{\textbf{Input:}}
	\renewcommand{\algorithmicensure}{\textbf{Output:}}
	\caption{Graph-Attention Multi-Agent Trust Region Algorithm}
	\label{alg1}
	\begin{algorithmic}[1]
            \STATE Initialize the joint policy ${\pi _0} = (\pi _0^1,...,\pi _0^n)$
            
            \STATE \textbf{for} \emph{k} = 0, 1, ..., \emph{K} \textbf{do} 
            
            \STATE \quad Encode the observation information $o(s,a)$ as a graph matrix ${\rm H} < \eta ,\zeta  > $ 
            
            \STATE \quad Compute attention value $Attention(q,\kappa ,\mu )$
            
            \STATE \quad Compute the advantage function: ${A_{{\pi _k}}}(Attention(q,\kappa ,\mu ))$
            
            \STATE \quad Compute $ Q = {\max _{s,a}}|{A_{{\pi _k}}}(s,a)|$ and $C = \frac{{4\gamma \varepsilon }}{{{{(1 - \gamma )}^2}}}$
            
            \STATE \quad Draw the arrangement ${i_{1:n}}$ of UAV BSs randomly 
            
            \STATE \quad \textbf{for} \emph{m} = 1 : \emph{n} \textbf{do}
            
            \STATE \quad\quad Update $\pi _{k + 1}^{{i_m}} = \arg {\max _{{\pi ^{{i_m}}}}}[L_{{\pi _k}}^{{i_{1:m}}}(\pi _{k + 1}^{{i_{1:m - 1}}},{\pi ^{{i_m}}}) - CD_{KL}^{\max }(\pi _k^{{i_m}},{\pi ^{i{}_m}})]$
            
            \STATE \quad \textbf{end for} 
            
            \STATE \textbf{end for}
	\end{algorithmic} 
\end{algorithm}

Figure 3 gives the framework of graph attention-based multi-agent trust region reinforcement learning. As shown in Figure 3, based on the actor-critic structure, the graph attention network is introduced to revise the critic network for the trust region MARL. Before the state information ${o_i}(s,a)$ of the intelligent agent is input into the critical network, it will be processed through the graph attention network in advance. Firstly, the connection information of intelligent UAV BSs and GUs are encoded into a graph matrix ${H}(\eta, \zeta)$. Then, the graph matrix and environmental information are assembled into the attention network ${Attention}(q,\kappa ,\mu )$ for feature extraction. Finally, they are input into the critical network for value evaluation.

\subsection{\textbf{Complexity Analysis}}
This paper compares the performance of the baseline MATR, the baseline IPPO \cite{49}, the baseline HATRPO \cite{45}, the MATR with single graph mechanism (Graph MATR), the MATR with single attention mechanism (Attention MATR), and the proposed Graph-Attention MATR. Meanwhile, the computational complexity of these algorithms are compared, which are summarized in Table I. The computational complexity mainly includes the complexity per layer, the input / output demensions of actor network, and the input / output demensions of the critic network.

\begin{table*}[htbp]
	\centering
	\caption{Comparison of computational complexity. \emph{n} is the sequence length, \emph{d} is the representation dimension, \emph{r} is the size of transition in self-attention, \emph{obs\_dim} is the dimension of observation information, and \emph{act\_dim} is the the dimension of action}
	\begin{tabular}{cccc}
		\toprule  
		 Algorithms & Complexity & Actor input / output demensions & Critic input / output demensions\\ 
		\midrule  
             MATR & $\emph{O}(\emph{n} \cdot \emph{d})$ &	$\emph{obs\_dim}\, /\,\emph{act\_dim}$ &	$\emph{obs\_dim} / 1$\\
             IPPO & $\emph{O}(\emph{d})$ &	$\emph{obs\_dim}\, /\,\emph{act\_dim}$ &	$\emph{obs\_dim} / 1$\\
             HATRPO & $\emph{O}(\emph{n} \cdot \emph{d})$ &	$\emph{obs\_dim}\, /\,\emph{act\_dim}$ &	$\emph{obs\_dim} / 1$\\
             Graph MATR & $\emph{O}(\emph{n}^{2} \cdot \ \emph{d})$ & $\emph{obs\_dim}\,/\,\emph{act\_dim}$ & $\emph{obs\_dim} / 1$\\
             Attention MATR & $\emph{O}(\emph{r} \cdot \emph{n} \ \cdot \ \emph{d})$ & $\emph{obs\_dim} / \emph{act\_dim}$ &	$\emph{r} \cdot \ \emph{obs\_dim} / 1$\\
             Graph-Attention MATR & $\emph{O}(\emph{r} \cdot  \emph{n}^{2} \cdot \ \emph{d})$ & $\emph{obs\_dim}\,/\,\emph{act\_dim}$ &	$\emph{r} \cdot \ \emph{obs\_dim} / 1$\\
		\bottomrule  
	\end{tabular}
\end{table*}

It can be observed that the difference of computational complexity between Graph-Attention MATR and other methods is not too significant. The computational complexity of the proposed algorithm in this paper is mainly determined by the size of transition, i.e. the number of GUs.

\subsection{\textbf{Equilibrium Solution for UAV-Assisted Communication Markov Game}}

This part explores the presence of NE for Markov game. 

\textbf{Theorem 1}: For the Markov game \emph{G} with \emph{M} UAV BSs, the optimal policy set $[{\Gamma ^*}({a_m},{a_{-m}}),m \in M]$ is a Nash equilibrium.

\textbf{Proof}: To begin, as stated in reference \cite{50}, the Markov game possesses a minimum of one NE.
After that, the algorithm's monotonic improvement has been formally demonstrated in Section III.B.2). Each agent is enabled to indefinitely converge towards the optimal strategy through the training procedure. Once the strategies converges to ${\Gamma ^*}$, all UAV BSs can adopt optimal policies during each game round, and no single agent can enhance profits by altering the policy independently. This meets the definition for Nash equilibrium.
When the following conditions hold:
\begin{eqnarray}
\begin{array}{l}
{\Gamma ^*} = \arg \mathop {\max }\limits_{{a_m},{a_{-m}}} {R_m},{R_{-m}},\, m \in M\\
\end{array}
\end{eqnarray}
${\Gamma ^*}({a_m},{a_{-m}})$ is a NE for the Markov game.

The proof is finished.

\textbf{Remark 1:} Reference \cite{49} proves that Markov game has at least one NE. But, it is a challenge to prove whether Nash equilibrium is unique, as well as to find the quantity of Nash equilibrium. In fact, checking for uniqueness of Nash equilibrium is NP-hard. In the practical training of algorithms, it will cost too much to calculate the optimal strategies. So, the agent will only conduct a limited episodes of training. Accordingly, we obtain the approximate solution of Nash equilibrium. This is where our work sets itself apart. 

\begin{figure}[!htb]
    \centering
    \subfloat[UAV BS 1]{\includegraphics[width=0.24\textwidth]{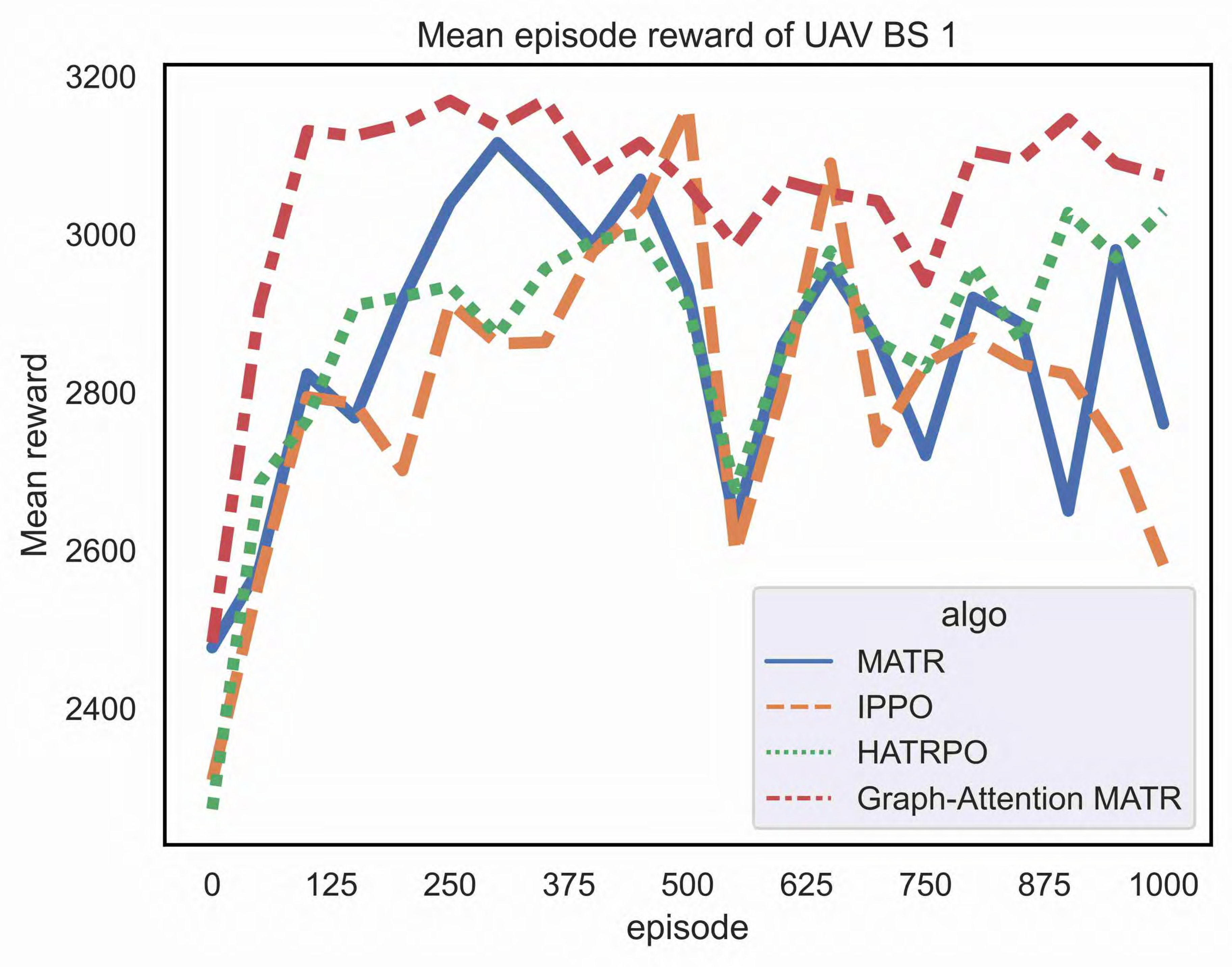}}
    \hfill
    \subfloat[UAV BS 2]{\includegraphics[width=0.24\textwidth]{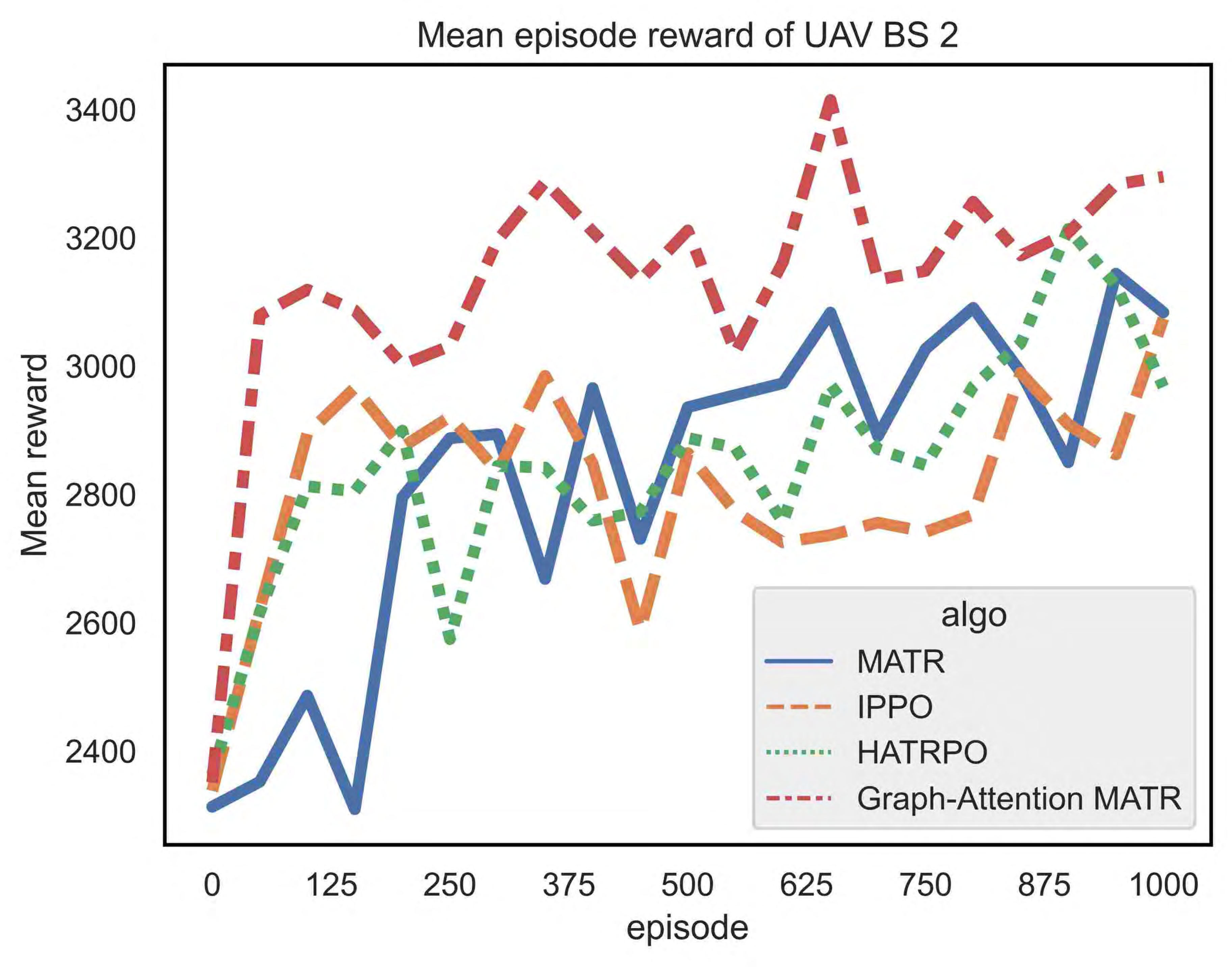}}
    \caption{Mean reward curves with two UAV BSs}
    \label{fig:two-subfigures-subfloat}
\end{figure}

\begin{figure}[htbp]
\centerline{\includegraphics[width=9cm]{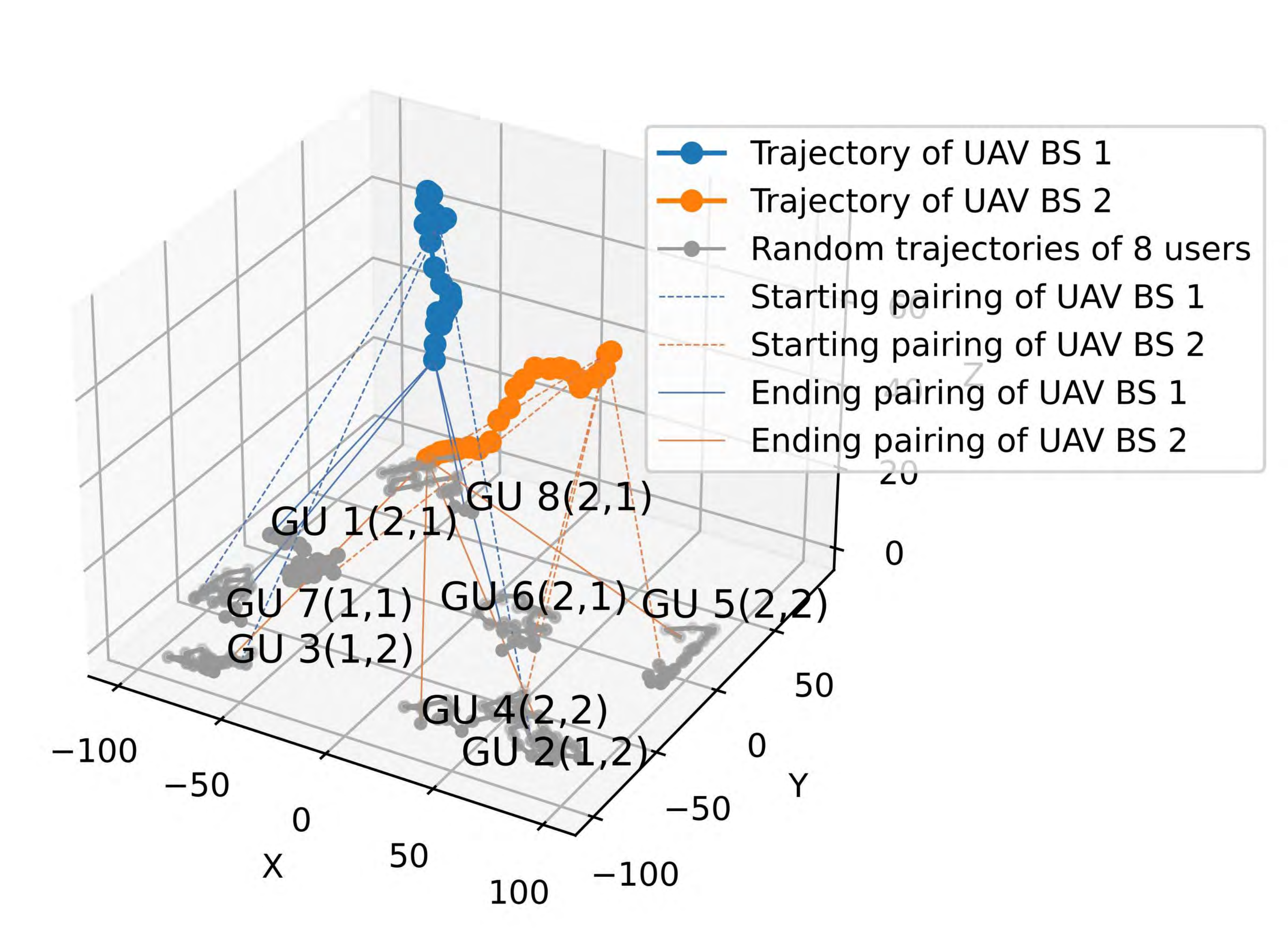}}
\caption{Trajectories and pairing with two UAV BSs. GU 1(2,1) means that GU 1 is paired with UAV BS 2 in the start and is paired with UAV BS 1 in the end, etc.}
\label{fig}
\end{figure}

\begin{figure}[!htb]
    \centering
    \subfloat[UAV BS 1]{\includegraphics[width=0.24\textwidth]{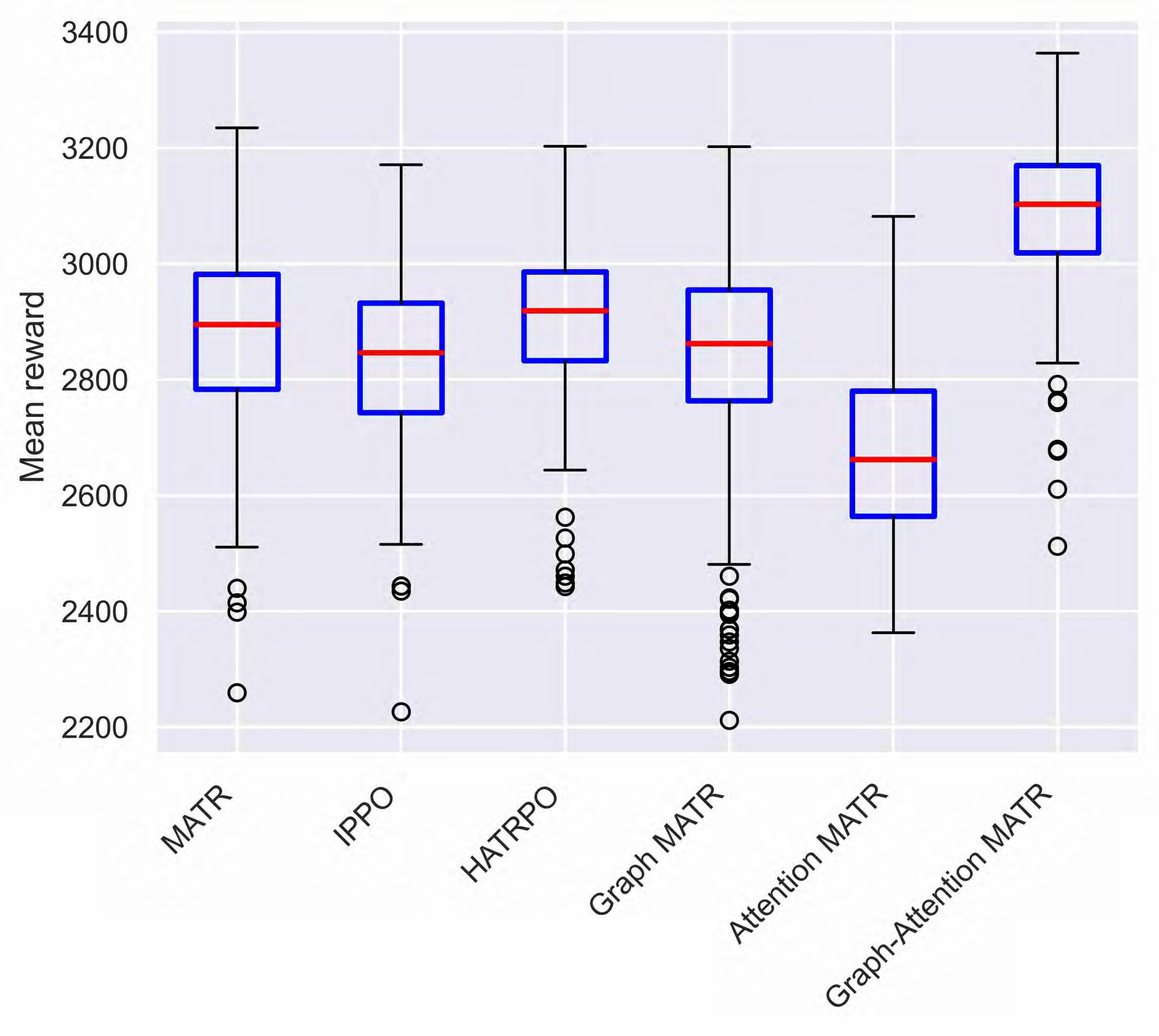}}
    \hfill
    \subfloat[UAV BS 2]{\includegraphics[width=0.24\textwidth]{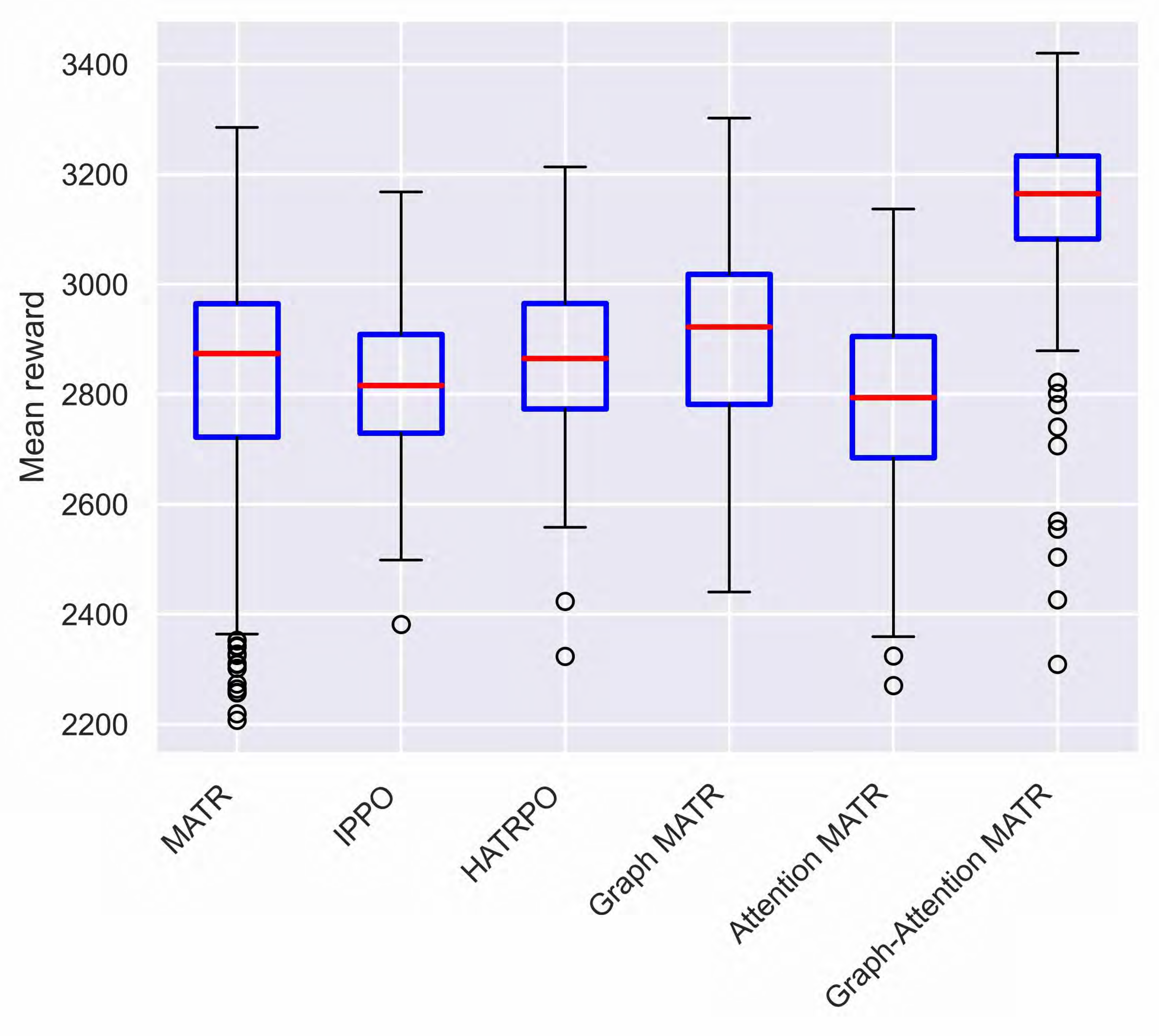}}
    \caption{Mean reward box plot with two UAV BSs}
    \label{fig:two-subfigures-subfloat}
\end{figure}

\begin{figure*}[!htb]
    \centering
    \subfloat[UAV BS 1]{\includegraphics[width=0.24\textwidth]{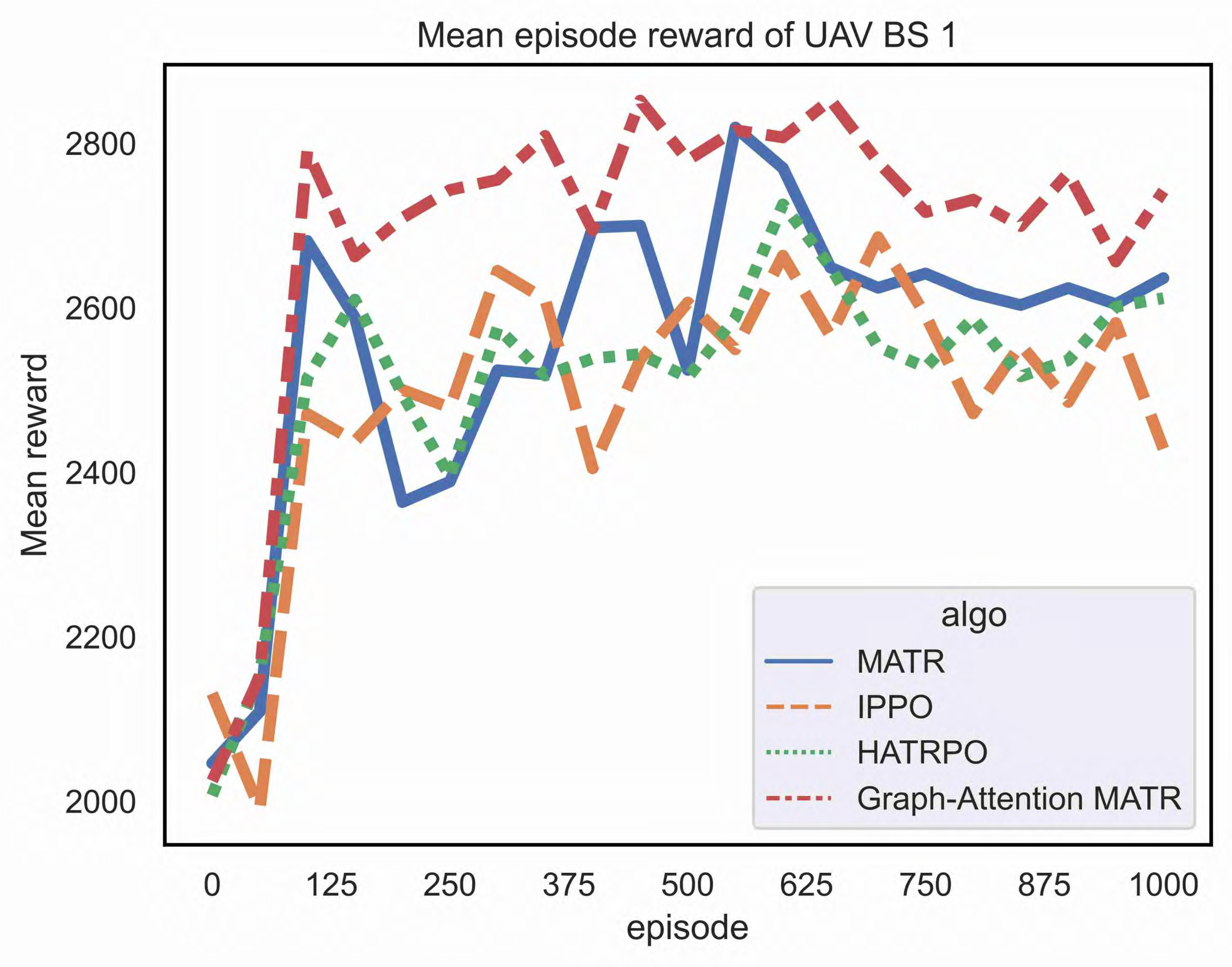}}
    \quad
    \subfloat[UAV BS 2]{\includegraphics[width=0.24\textwidth]{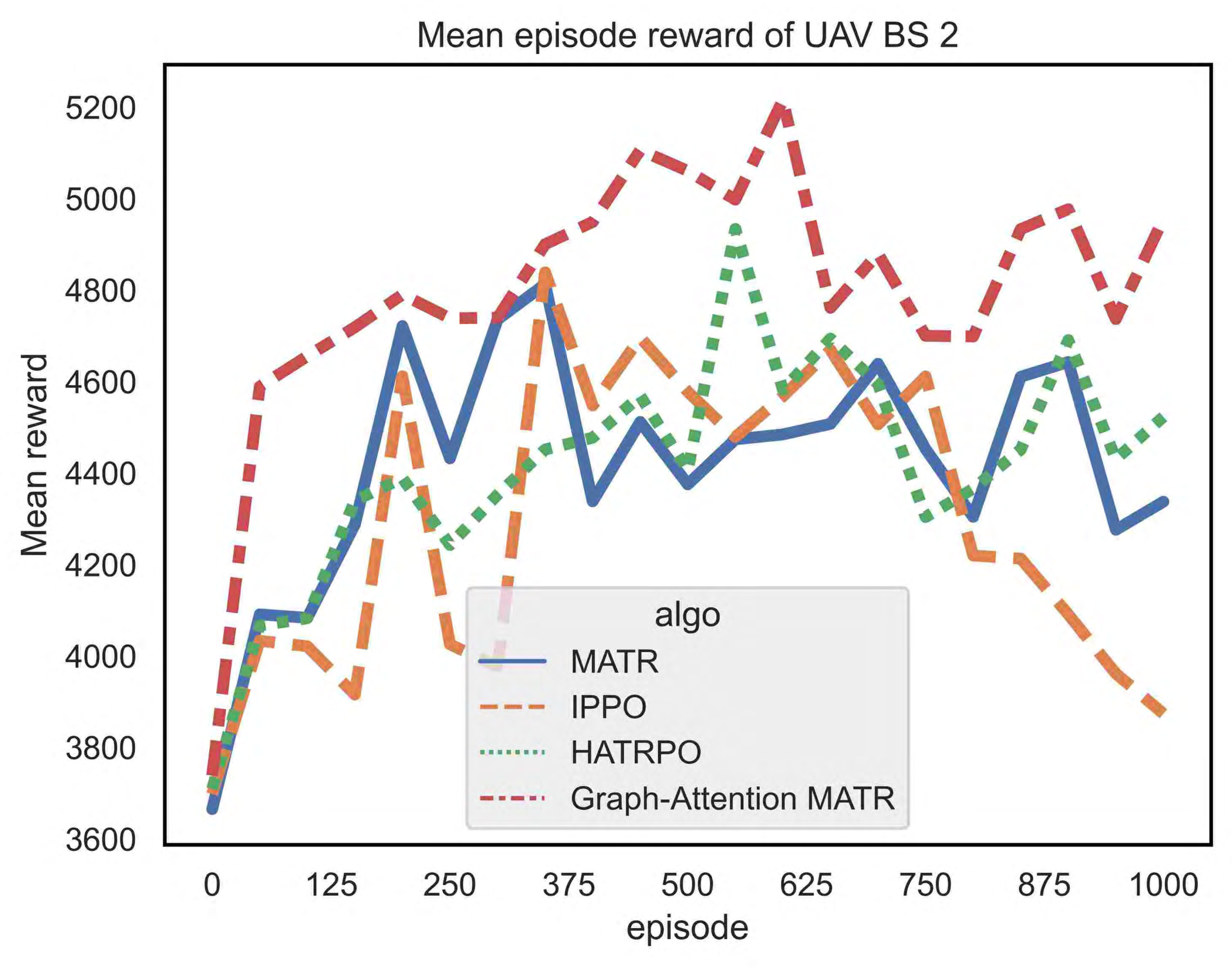}}
    \quad
    \subfloat[UAV BS 3]{\includegraphics[width=0.24\textwidth]{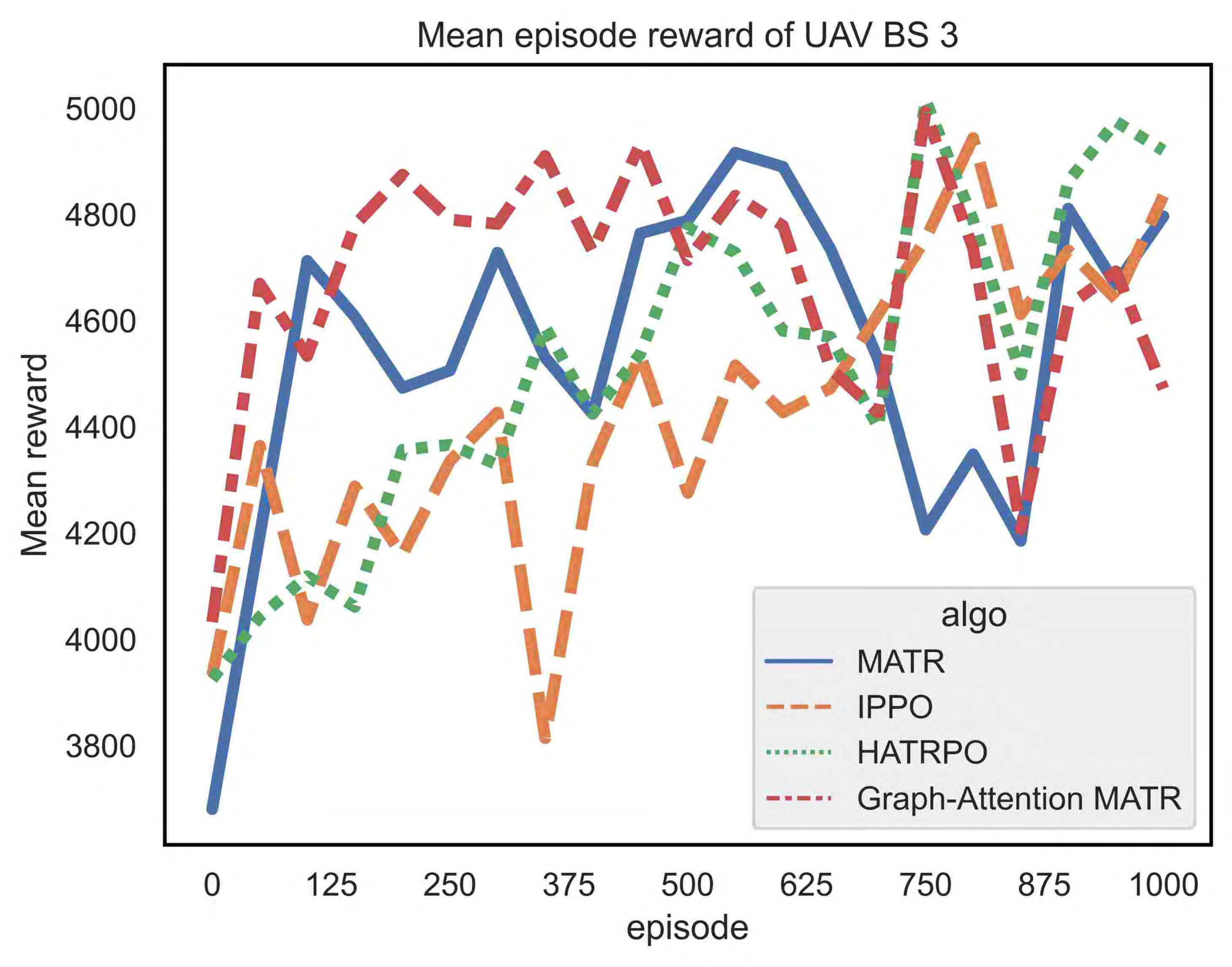}}
    \caption{Mean reward curves with three UAV BSs}
    \label{fig:two-subfigures-subfloat}
\end{figure*}

\begin{figure}[htbp]
\centerline{\includegraphics[width=7.5cm]{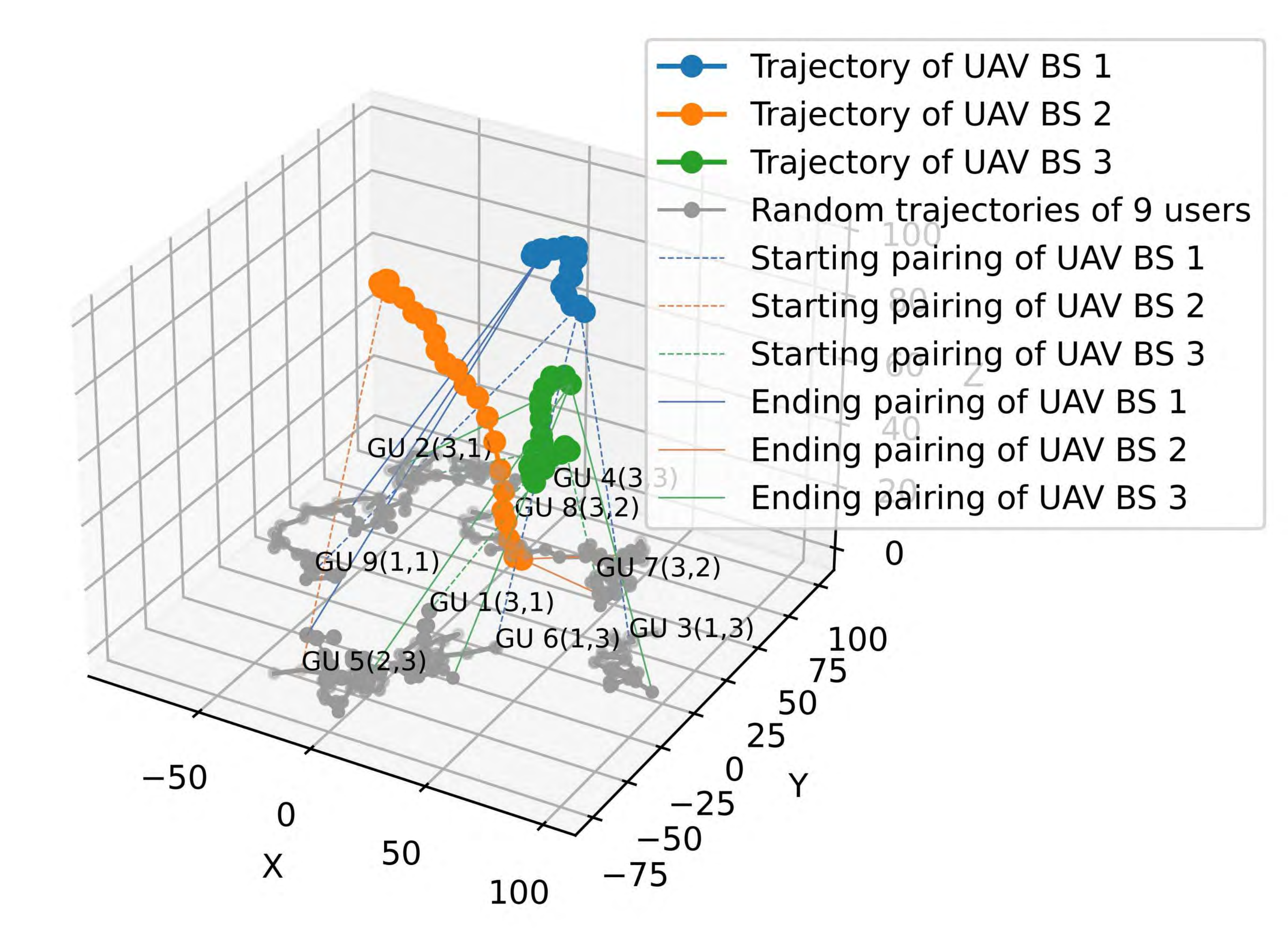}}
\caption{Trajectories and pairing with three UAV BSs
}
\label{fig}
\end{figure}

\begin{figure*}[!htb]
    \centering
    \subfloat[UAV BS 1]{\includegraphics[width=0.25\textwidth]{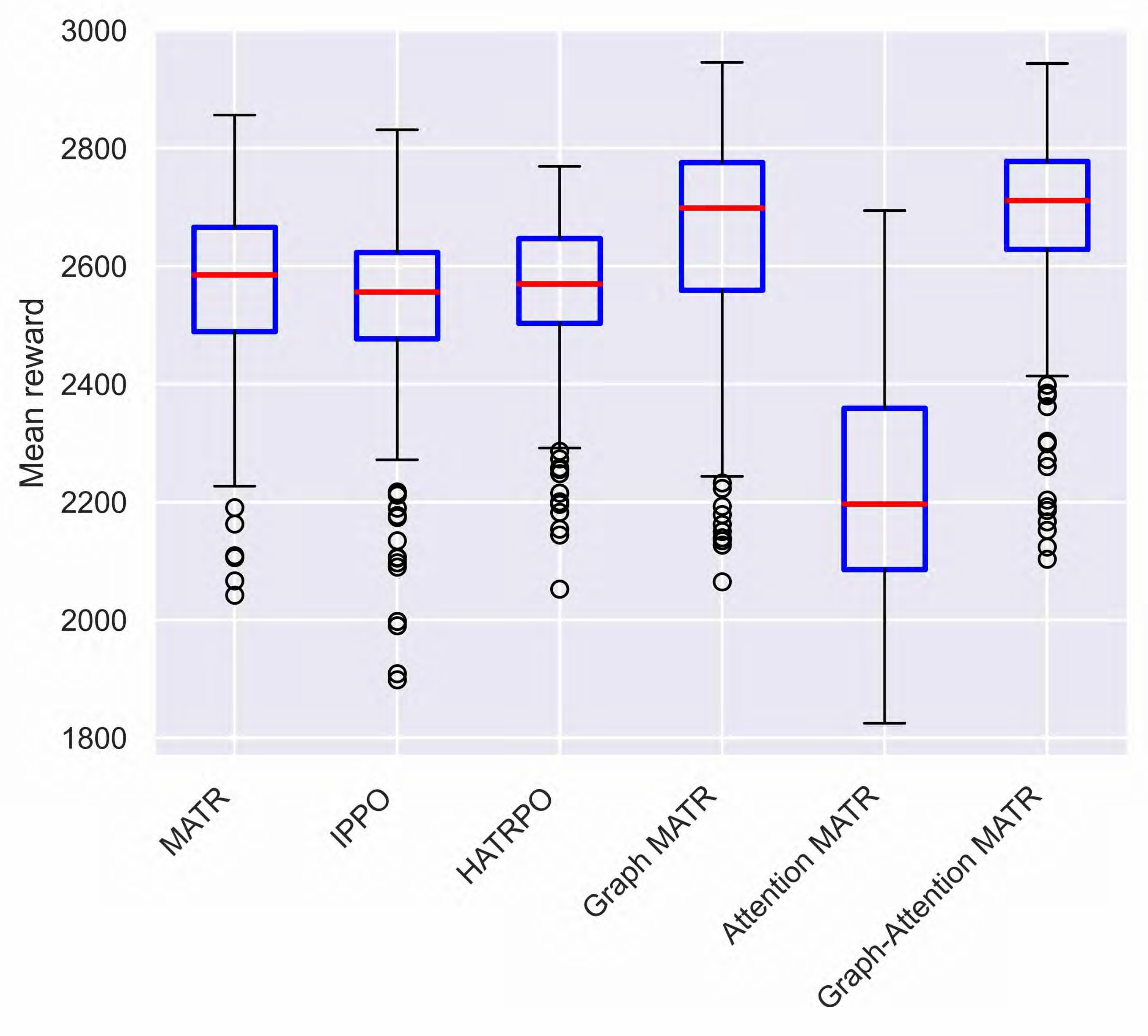}}
    \quad
    \subfloat[UAV BS 2]{\includegraphics[width=0.25\textwidth]{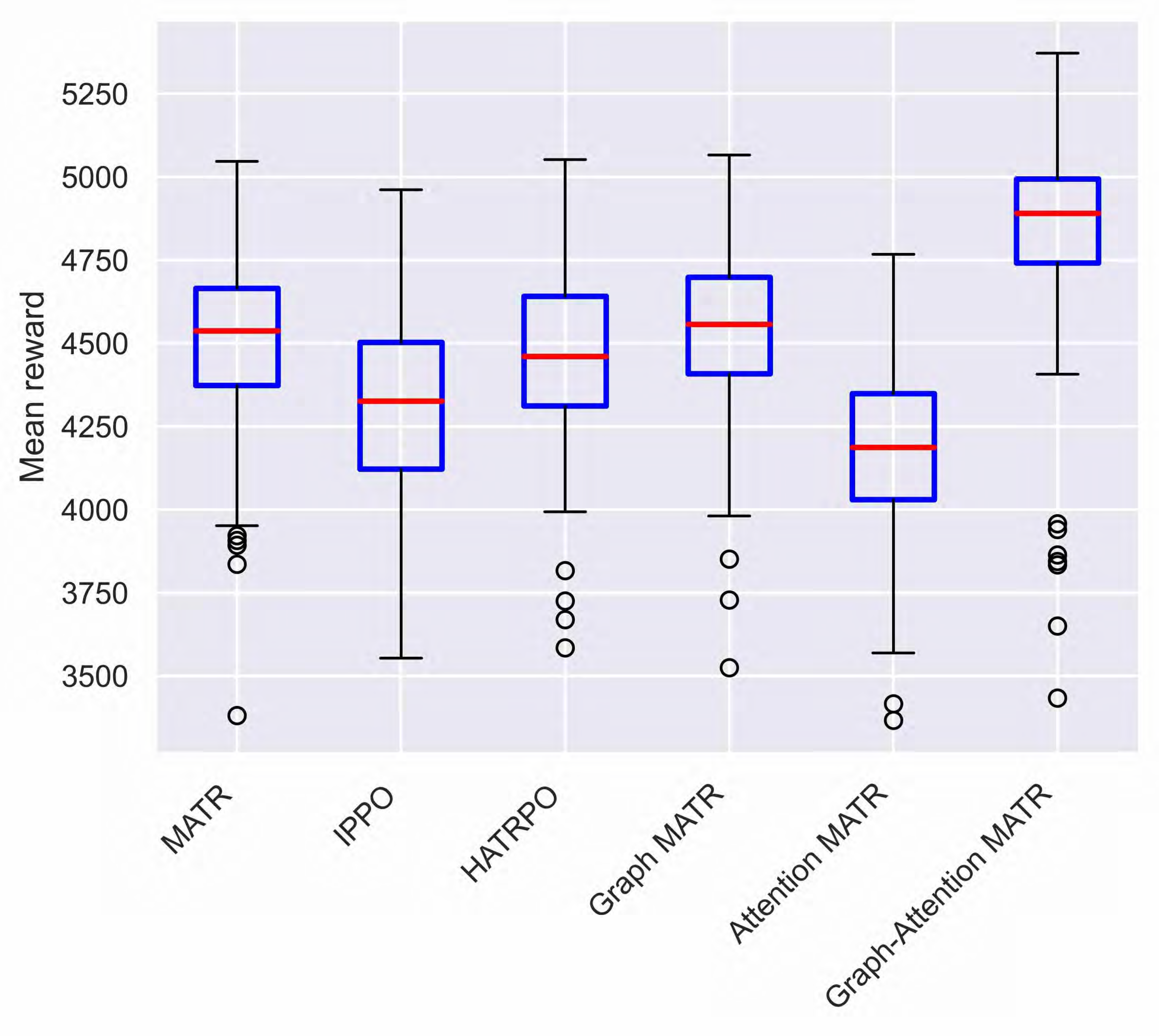}}
    \quad
    \subfloat[UAV BS 3]{\includegraphics[width=0.25\textwidth]{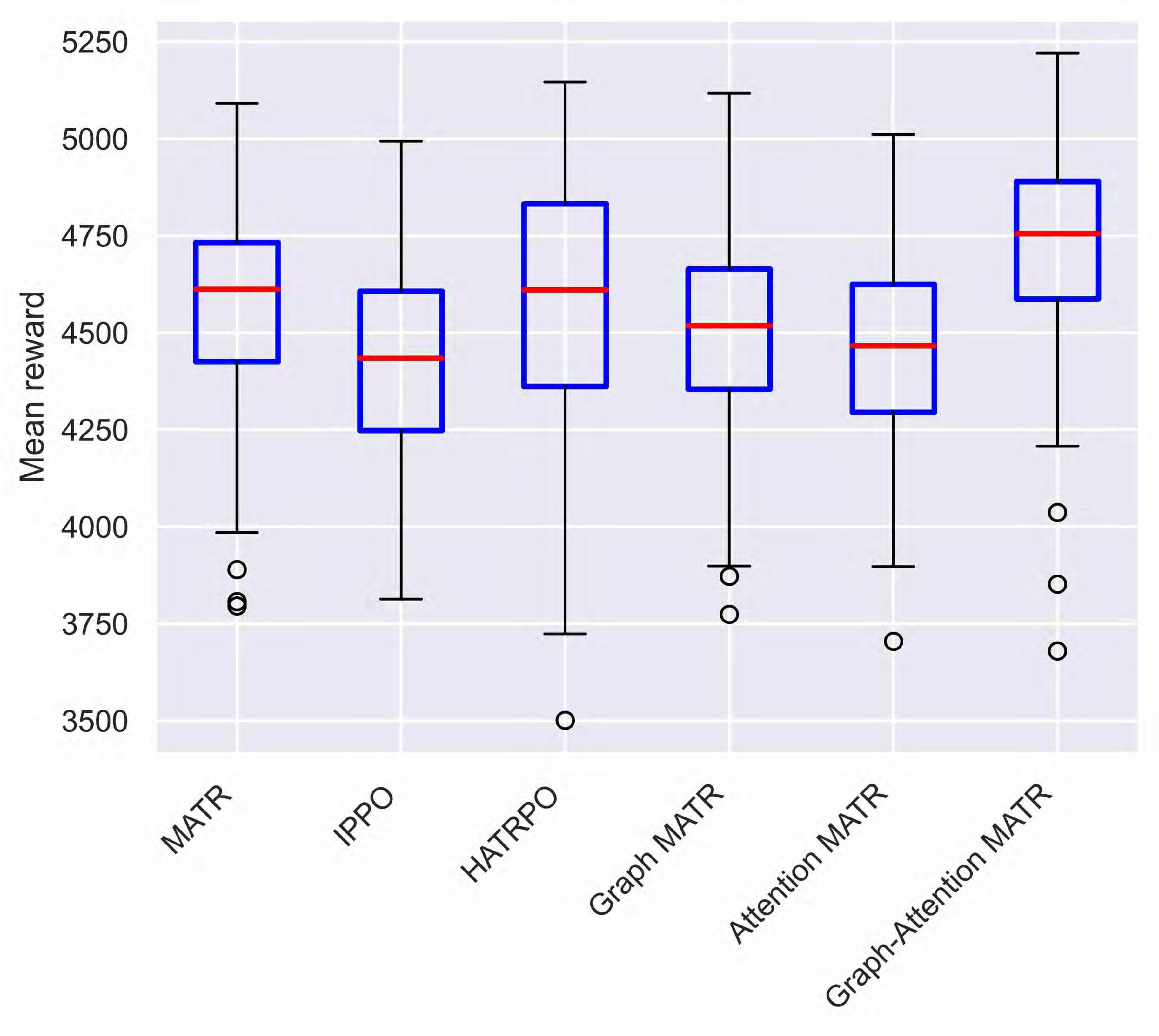}}
    \caption{Mean reward box plot with three UAV BSs}
    \label{fig:two-subfigures-subfloat}
\end{figure*}

\begin{figure*}[!htb]
    \centering
    \subfloat[UAV BS 1]{\includegraphics[width=0.235\textwidth]{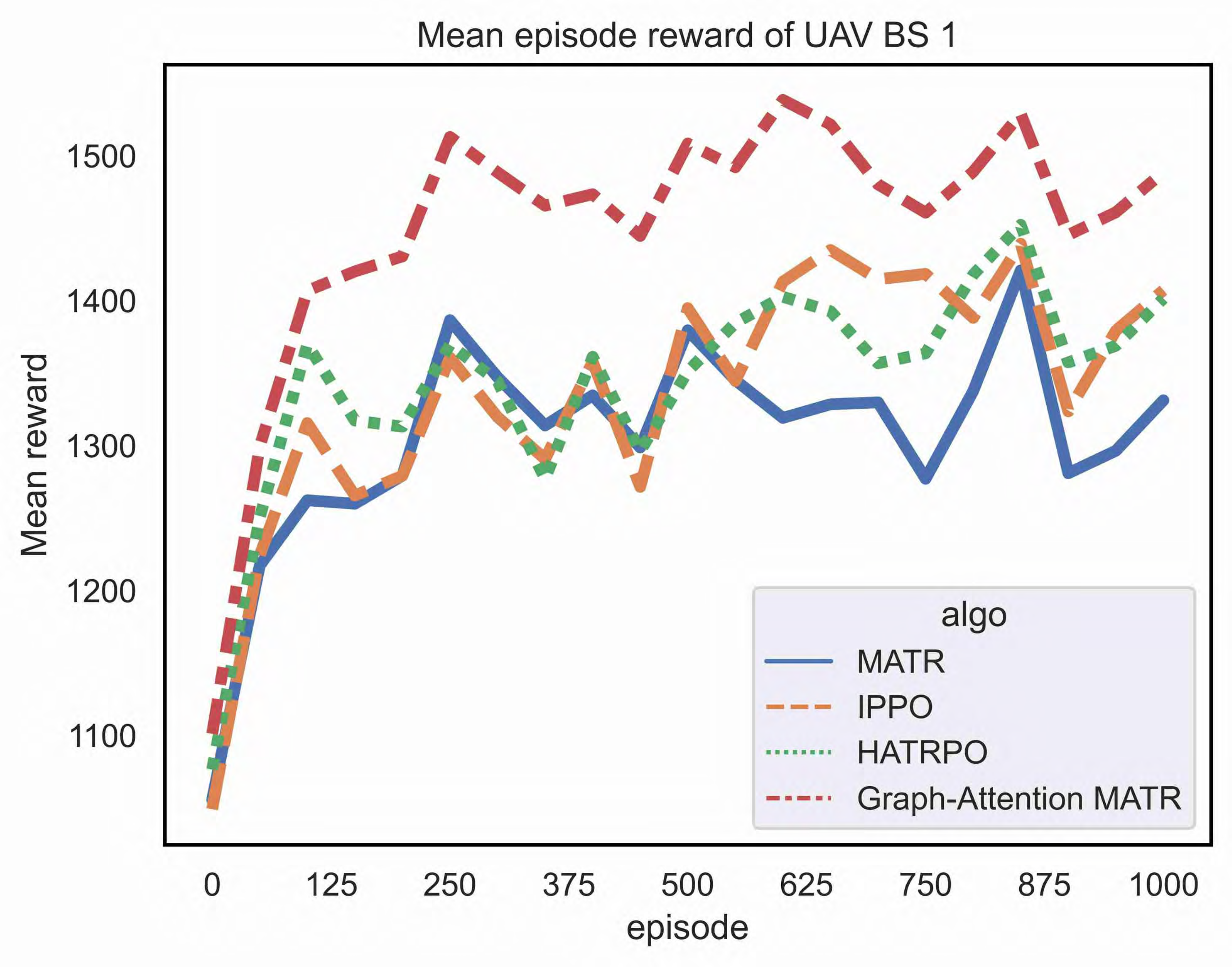}}
    \quad
    \subfloat[UAV BS 2]{\includegraphics[width=0.235\textwidth]{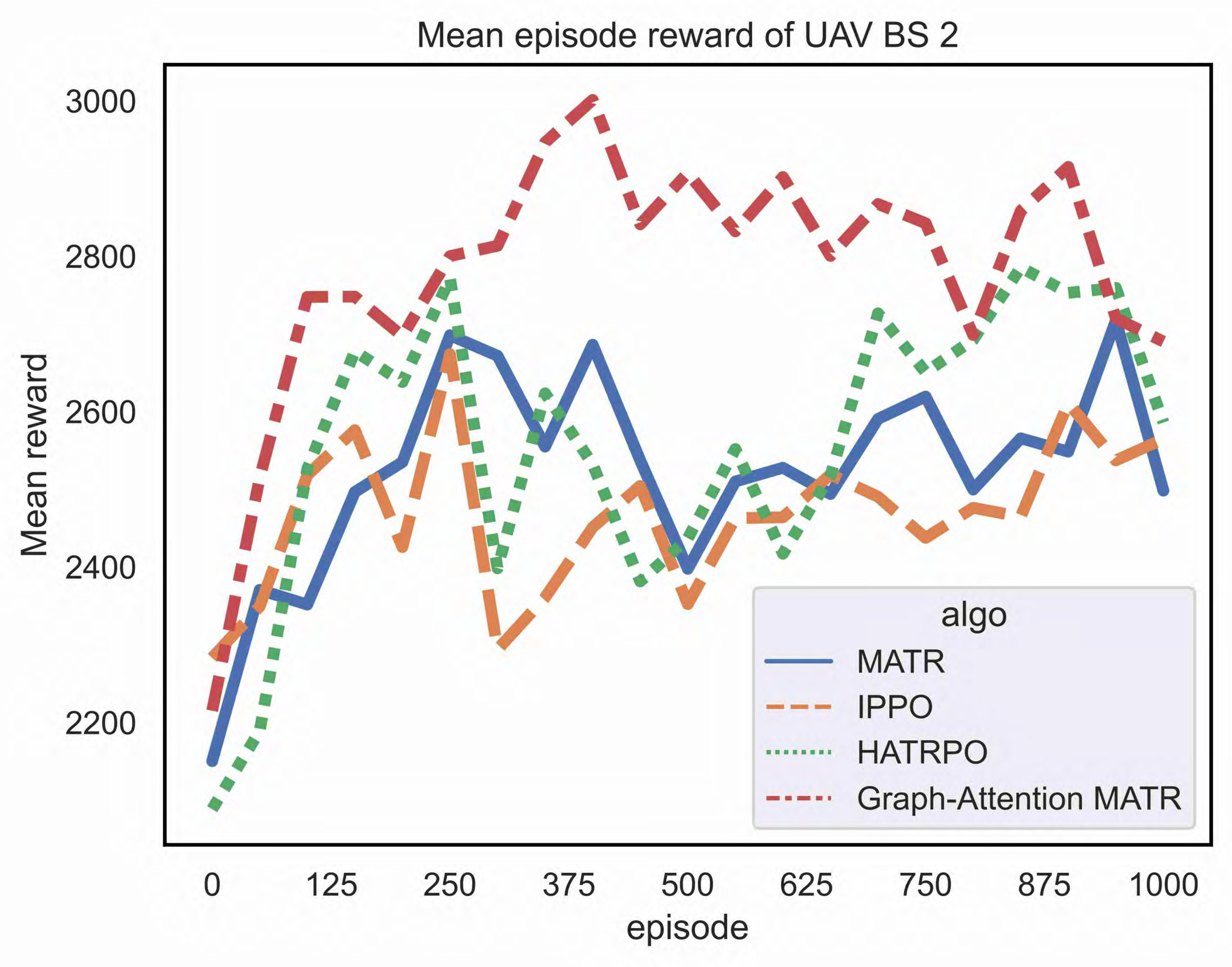}}
    \quad
    \subfloat[UAV BS 3]{\includegraphics[width=0.235\textwidth]{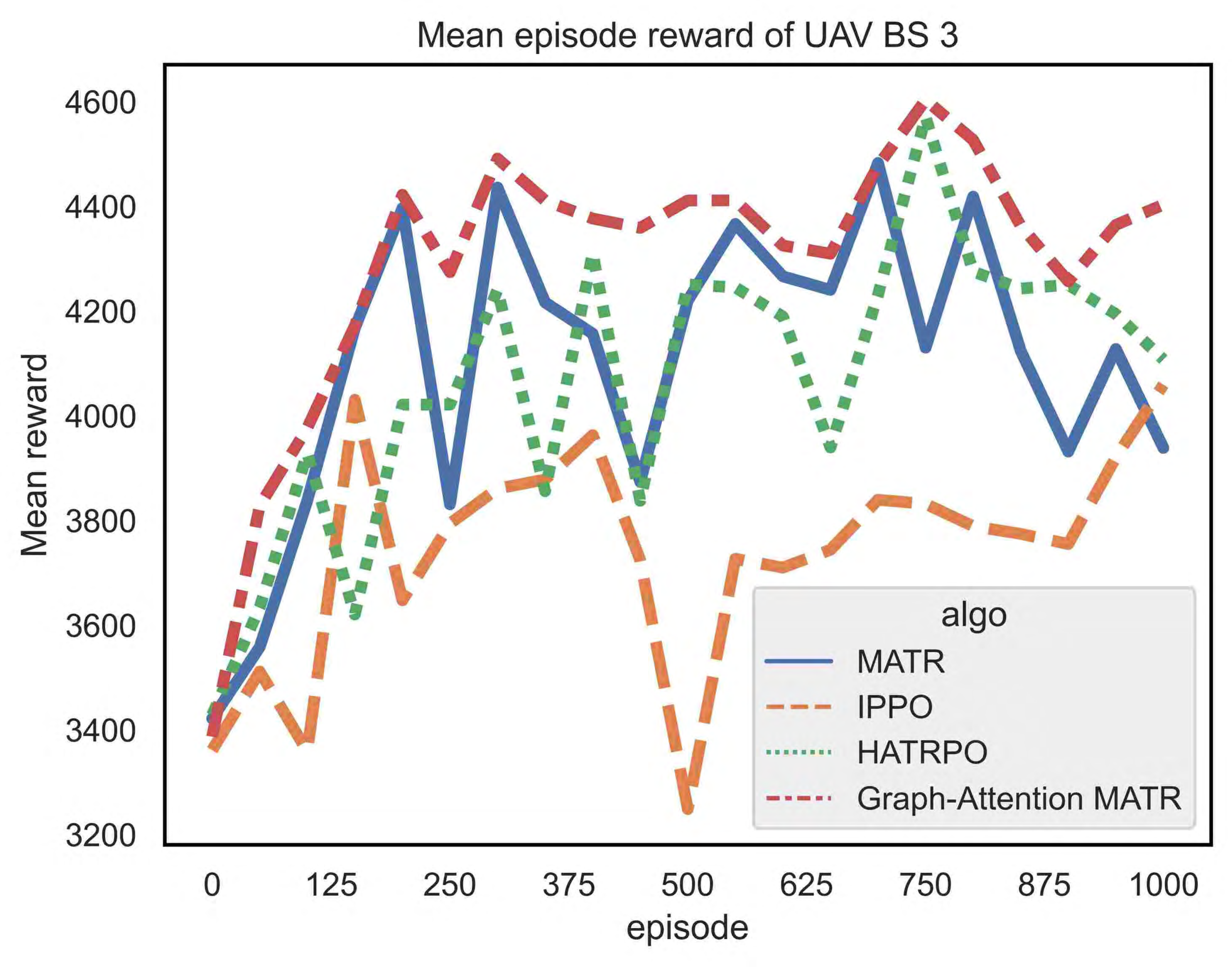}}
    \quad
    \subfloat[UAV BS 4]{\includegraphics[width=0.235\textwidth]{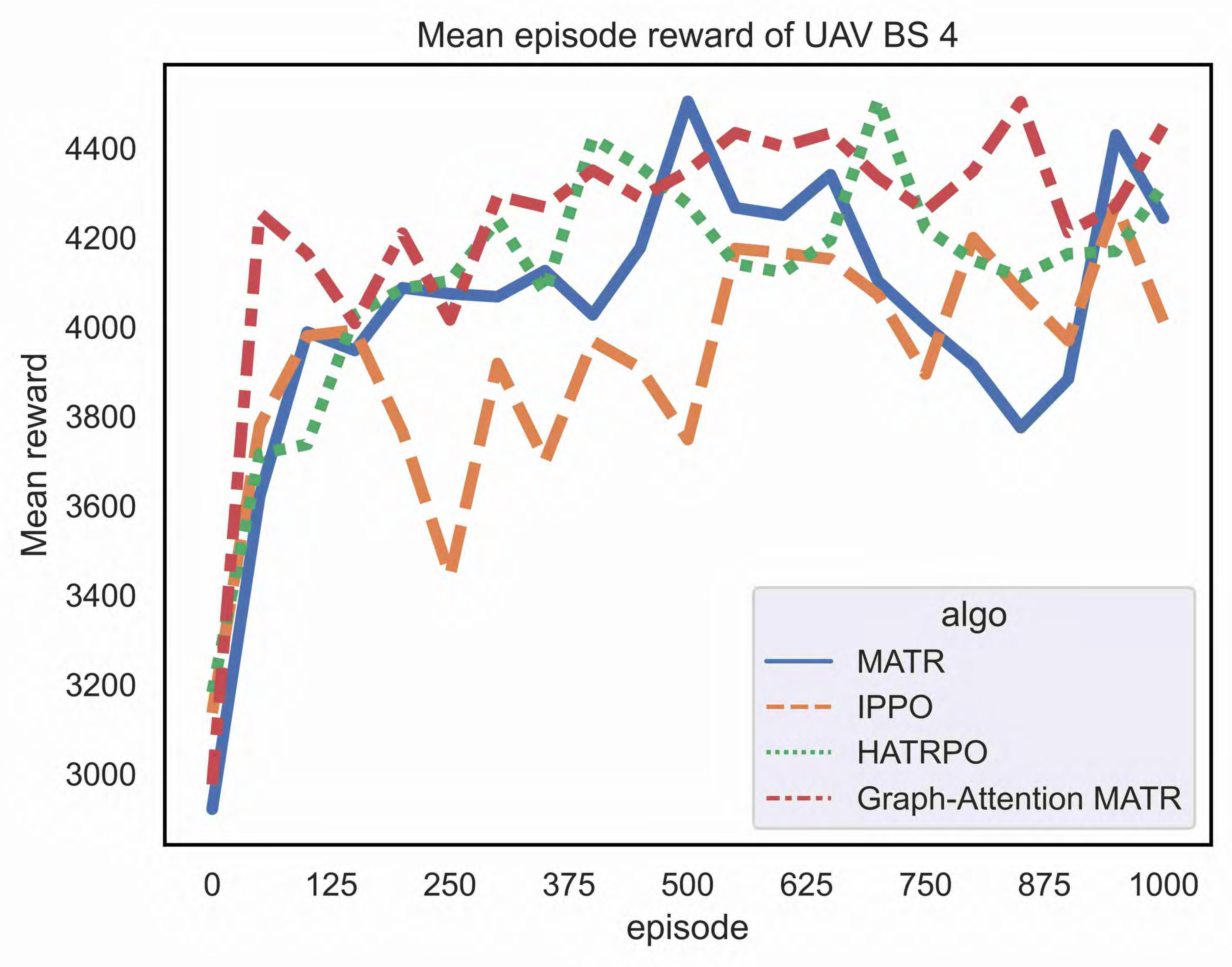}}
    \quad
    \caption{Mean reward curves with four UAV BSs}
\end{figure*}

\begin{figure*}[!htb]
    \centering
    \subfloat[UAV BS 1]{\includegraphics[width=0.235\textwidth]{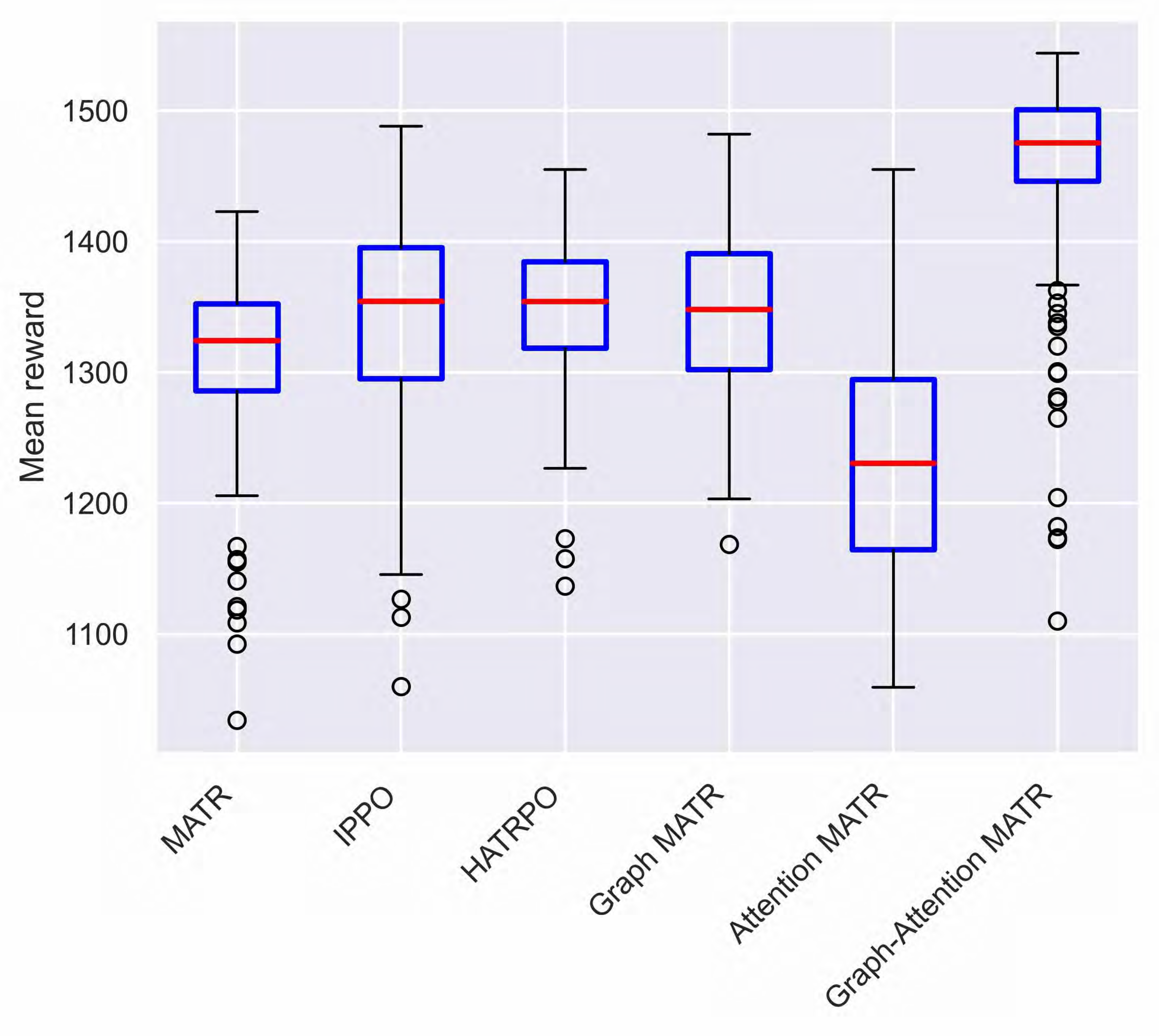}}
    \quad
    \subfloat[UAV BS 2]{\includegraphics[width=0.235\textwidth]{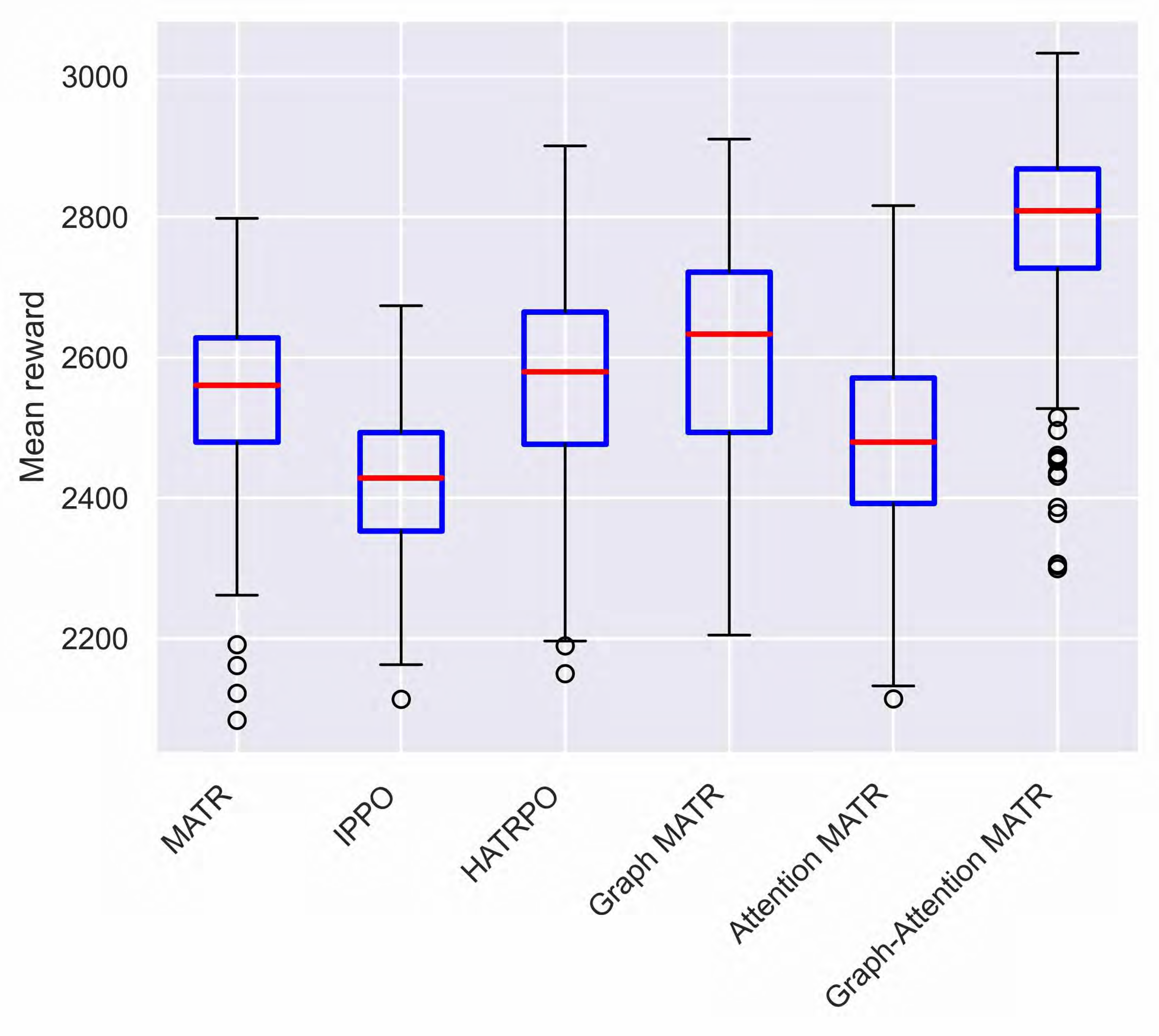}}
    \quad
    \subfloat[UAV BS 3]{\includegraphics[width=0.235\textwidth]{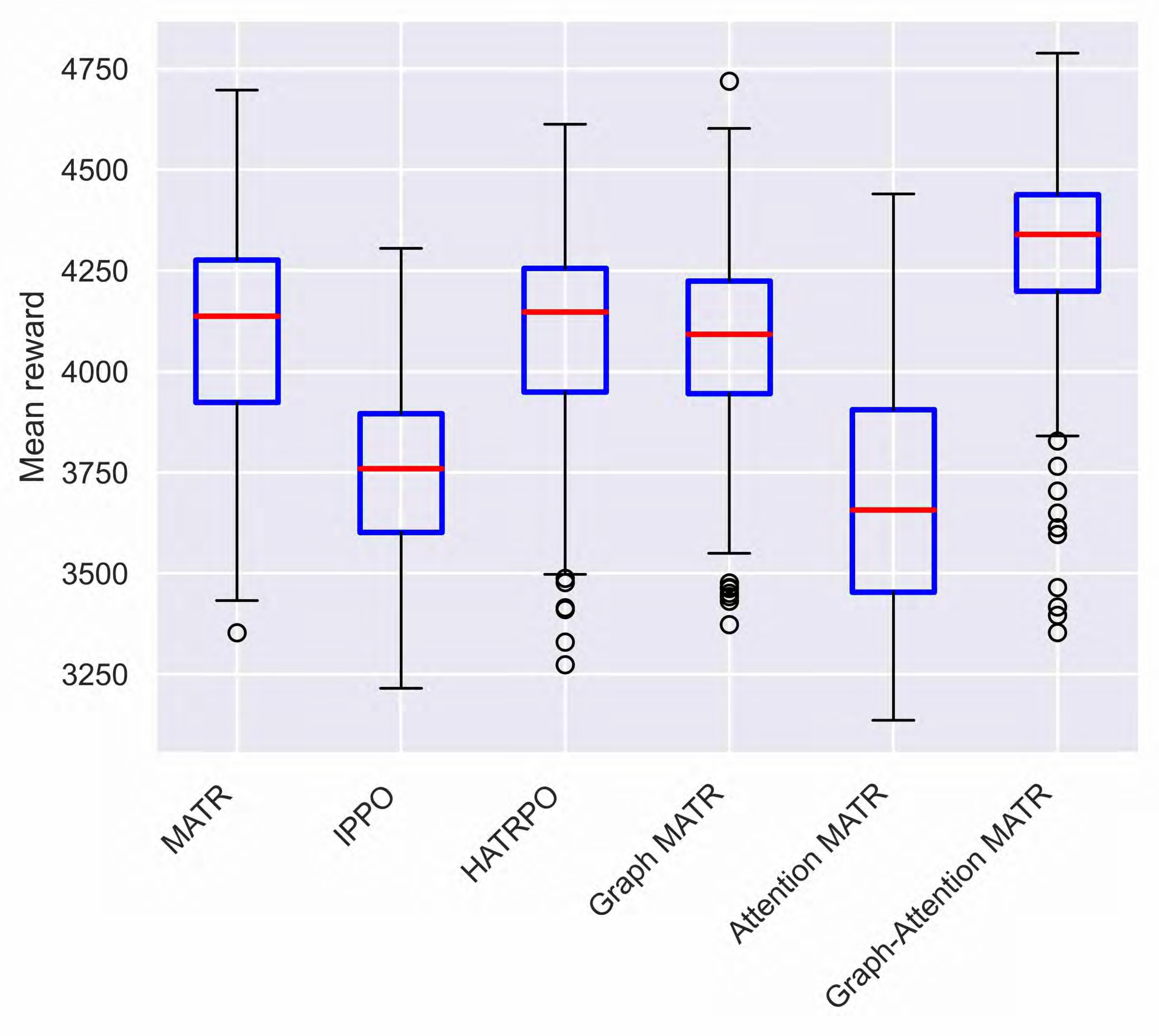}}
    \quad
    \subfloat[UAV BS 4]{\includegraphics[width=0.235\textwidth]{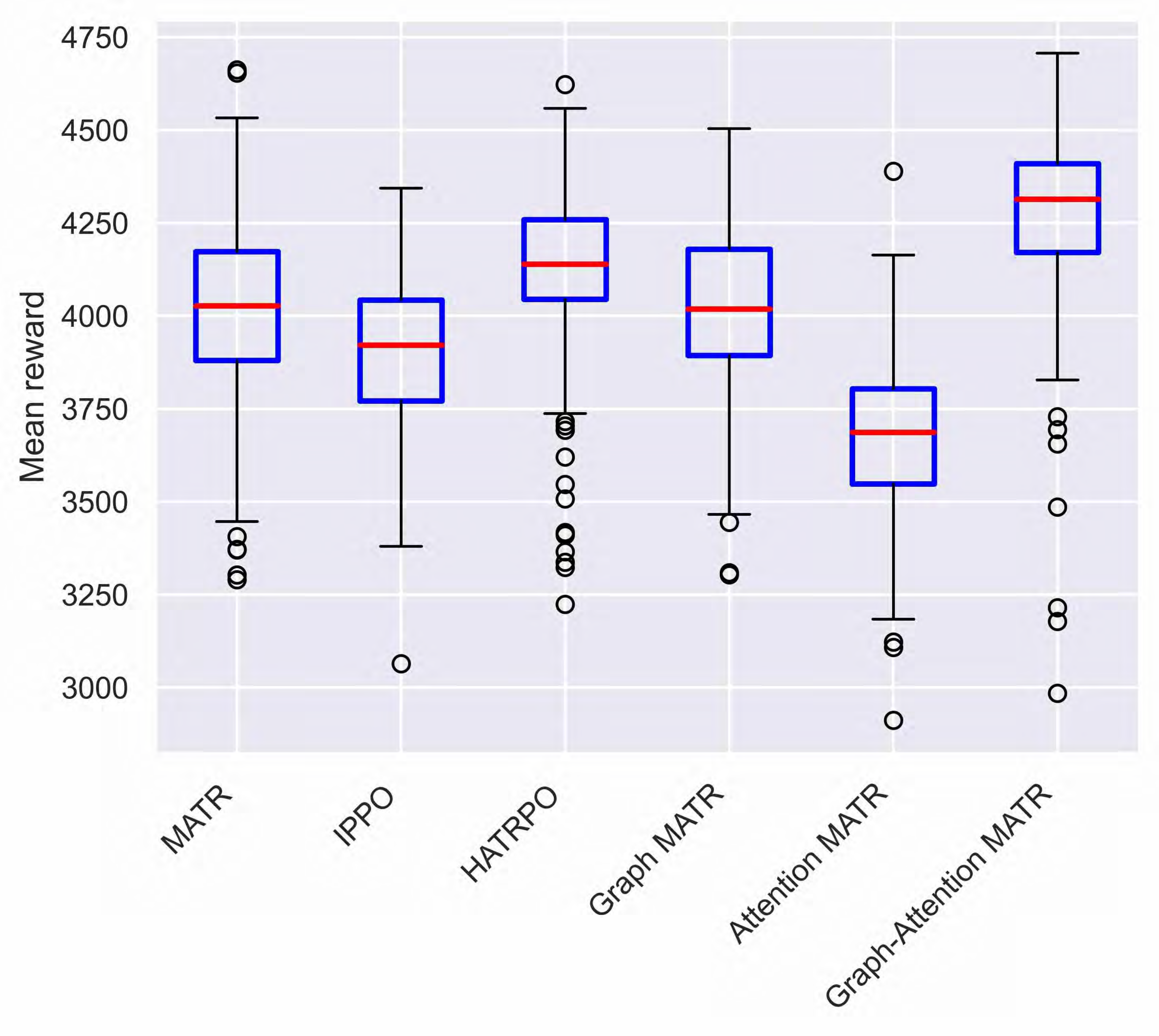}}
    \quad
    \caption{Mean reward box plot with four UAV BSs}
\end{figure*}

\begin{figure}[htbp]
\centerline{\includegraphics[width=9.5cm]{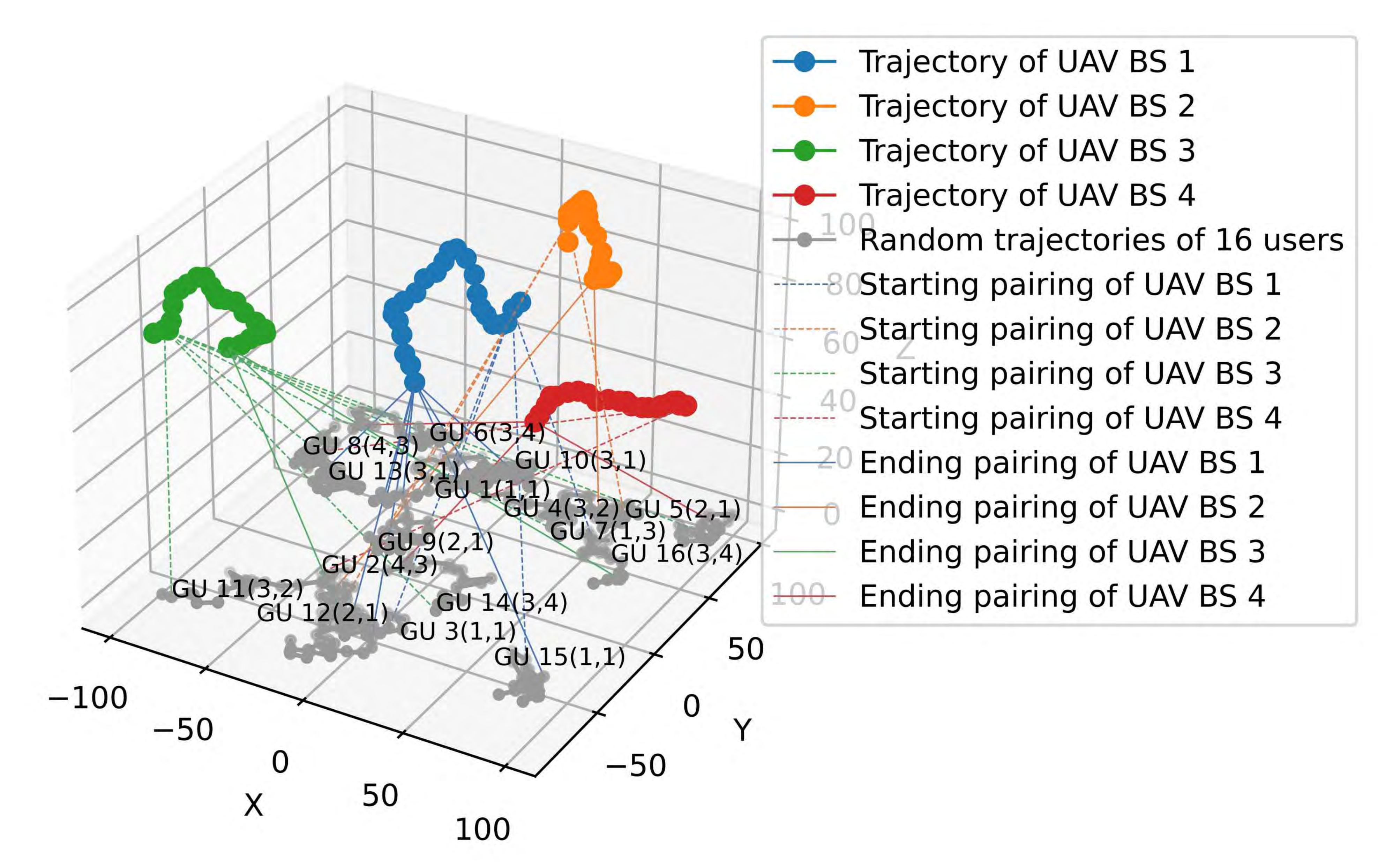}}
\caption{Trajectories and pairing with four UAV BSs
}
\label{fig}
\end{figure}
\begin{table*}[htbp]
	\centering
	\caption{Final episode reward with 2 UAV BSs}
	\begin{tabular}{ccccc}
		\toprule  
		   & UAV BS 1 & UAV BS 2 \\ 
		\midrule  
             MATR & 2760.250 & 3083.469\\
             IPPO & 2581.505 & 3074.054\\
             HATRPO & 3029.567 & 2967.351\\
             Graph MATR & 2876.115 & 3046.190\\
             Attention MATR & 2788.917 & 2819.708\\
             Graph-Attention MATR (This paper) & \textbf{3073.919} & \textbf{3295.174}\\
		\bottomrule  
	\end{tabular}
\end{table*}

\begin{table*}[htbp]
	\centering
	\caption{Final episode reward with 3 UAV BSs}
	\begin{tabular}{ccccc}
		\toprule  
		   & UAV BS 1 & UAV BS 2 & UAV BS 3\\ 
		\midrule  
             MATR & 2636.052 & 4337.268 & 4796.336\\
             IPPO & 2428.499 & 3873.794 & 4835.297\\
             HATRPO & 2611.63 & 4522.892 & \textbf{4920.298}\\
             Graph MATR & 2623.020 & 4584.723 & 4654.053\\
             Attention MATR & 2172.944 & 3978.846 & 4160.447\\
             Graph-Attention MATR (This paper) & \textbf{2741.980} & \textbf{4955.742} & 4471.564\\
		\bottomrule  
	\end{tabular}
\end{table*}

\begin{table*}[htbp]
	\centering
	\caption{Final episode reward with 4 UAV BSs}
	\begin{tabular}{ccccc}
		\toprule  
		   & UAV BS 1 & UAV BS 2 & UAV BS 3 & UAV BS 4\\ 
		\midrule  
             MATR & 1331.177 & 2498.194 & 3938.529 & 4243.466\\
             IPPO & 1407.159 & 2564.732 & 4059.404 & 4012.171\\
             HATRPO & 1401.452 & 2585.839 & 4107.441 & 4316.122\\
             Graph MATR & 1397.712 & 2592.799 & 4252.433 & 4142.618\\
             Attention MATR & 1166.262 & 2506.635 & 3262.909 & 3783.110\\
             Graph-Attention MATR (This paper) & \textbf{1488.718} & \textbf{2690.066} & \textbf{4403.375} & \textbf{4449.627}\\
		\bottomrule  
	\end{tabular}
\end{table*}
\section{Performance Evaluation}
This part gives the simulation results with different number of UAV BSs and GUs.

To validate the effectiveness of the proposed graph-attention MATR algorithm, five baseline methods are compared as: (1) the baseline MATR; (2) the baseline IPPO; (3) the baseline HATRPO; (4) the Graph MATR; (5) the Attention MATR.

In this paper, Adam optimizer is adopted to update the neural network \cite{51}. The following provides a detailed analysis.
\subsection{\textbf{Parameter Settings}}
In this paper, the area \emph{D} is set as a cuboid space of 200*200*100 \emph{m}, and a Cartesian coordinate system is constructed based on this. The GU moves on the 2D ground of (-100, 100)*(-100, 100). The UAV BS flies in the 3D space. For safety, the flight height is set to (10 , 100). At the beginning of each episode, all UAV BSs take off from random positions, and each GU moves randomly at an interval speed of [0m/s, 10m/s]. In all experiments, each UAV BS has rated transmit power ${P_{total}} = 10dBm$, shared total bandwidth ${B_{total}}$ is 30MHz, communication carrier frequency is ${f_c} = 2GHz$, noise power spectral density is ${n_0} = {10^{ - 17}}W/Hz$, and minimum divisible bandwidth is ${b_{\min }} = 0.1MHz$. The service duration of each round is assumed to be ${T_{\max }} = 1000s$, and each decision interval is 1 s.
\subsection{\textbf{Result Analysis}}
In this part, the results of the graph attention-based trust region MARL approach with different scenes are discussed.

1) \textbf{Results with two UAV BSs and eight GUs}

As shown in Figure 4, ablation experiments were conducted to compare the proposed algorithm and other baseline algorithms. As shown in the mean episode reward curve, the convergence of the graph-attention MATR is the best. The overall mean episode reward obtained is superior to the other three methods. Table II provides the final episode reward of two UAV BSs based on different algorithms. It can be seen that among the 1000 episodes, based on the algorithm proposed in this article, all two UAV BSs achieved the highest cumulative reward. This validates the superior convergence performance of the proposed method from another aspect. Figure 5 gives the trajectories, starting pairing, and ending pairing of last 25 steps for each UAV BS in the last episode. Starting from the initial position, UAV BSs make decision based on the experience learned from historical data, changing their spatial positions and pairing policies to achieve optimal communication performance. At the beginning, UAV BS 1 chooses to pair with GU 2, GU 3, and GU 7; UAV BS 2 chooses to pair with GU 1, GU 4, GU 5, GU 6, and GU 8. At the end, UAV BS 1 adjusts its policy to pair with GU 1, GU 6, GU 7, and GU 8; UAV BS 2 adjusts its policy to pair with GU 2, GU 3, GU 4, and GU 5. Figure 6 gives the mean reward box plot of two UAV BSs under four algorithms. It can also be seen that, compared to other benchmark methods, the proposed algorithm has complete advantages in terms of mean reward distribution.

2) \textbf{Results with three UAV BSs and nine GUs} 

As shown in Figure 7, ablation experiments were conducted to compare the proposed algorithm and other baseline algorithms. As shown in the mean episode reward curve, the convergence effect of the proposed graph-attention MATR is above other curves in general. The overall mean episode reward obtained is superior to the other three methods. Table III provides the final episode reward of three UAV BSs based on different algorithms. It can be seen that among the 1000 episodes, based on the algorithm proposed in this article, two out of three UAV BSs achieved the highest cumulative rewards. This validates the superior convergence performance of the proposed method from another aspect. Figure 8 gives the trajectories, starting pairing, and ending pairing of last 25 steps for each UAV BS in the last episode. Starting from the initial position, UAV BSs make decision based on the experience learned from historical data, changing their spatial positions and pairing policies to achieve optimal communication performance. Figure 9 gives the mean reward box plot of three UAV BSs under four algorithms. It can also be seen that, compared to other benchmark methods, the proposed algorithm has advantages at the highest median.

3) \textbf{Results with four UAV BSs and sixteen GUs}

In Figure 10, ablation experiments were carried out to compare the proposed algorithm with other baseline algorithms. As shown in the mean episode reward curve, the convergence effect of the proposed algorithm is generally above other curves. Specifically, the overall mean episode reward obtained is superior to the other three methods. Table IV provides the final episode reward based on different algorithms. It can be seen that among the 1000 episodes, based on the algorithm proposed in this paper, all four UAV BSs achieved the highest cumulative reward. This validates the superior convergence performance of the proposed method from another aspect. Figure 11 gives the mean reward box plot of four UAV BSs with four algorithms. It can also be seen that, compared to other benchmark methods, the proposed algorithm has advantages in terms of mean reward distribution. Figure 12 gives the trajectories, starting pairing, and ending pairing of last 25 steps for every UAV BS in the last episode. Starting from the initial position, UAV BSs make decision based on the experience learned from historical data, changing their spatial positions and pairing policies to achieve optimal performance.

\textbf{Remark 2:} To sum up, in different scenarios, the proposed graph attention multi-agent trust region reinforcement learning significantly improves the convergence performance, which supports good solution for the trajectory design and resource allocation in multi-UAV assisted communication Markov game. This is because that trust region MARL guarantees the monotonic convergence firstly. Secondly, graph recurrent network extracts useful information and patterns from large amount of state information. Thirdly, the attention mechanism provides additional information transmission and weighting, so that the critic network can more accurately evaluate the behavior value for UAV BSs. It provides more reliable feedback signals and helping the actor network to update the strategy more effectively.

\section{Conclutions}
To deal with the dynamic trajectory design and resource assignment in multi-UAV assisted communication, Markov game is used to model the communication network. Then, a novel graph attention-based multi-agent trust region reinforcement learning framework is proposed to optimize pairing strategies with GUs, mobile trajectories, power allocation strategies, and bandwidth allocation strategies for all UAV BSs. Graph recurrent network is introduced to process and analyze complex topology and extract useful patterns from interactive data. The attention mechanism is designed to provide additional weighting for conveyed information between UAV BSs, so that the critic network of each UAV BS can more accurately evaluate the value of behavior. It provides more reliable feedback signals and helps the UAV BS update parameters of the actor network more effectively. This also allows UAV BSs to learn the moving policies and communication strategies of others, thereby optimizing their own policy. The introduced algorithm exhibits favorable performance and superior convergence compared to baseline strategies. Equally important is that, well-trained intelligent UAV BSs can attain stable polices, thereby forming an approximate Nash equilibrium for the downlink communication Markov game.


\end{document}